\documentclass[notitlepage,superscriptaddress,nofootinbib,tightenlines,twocolumn]{revtex4-1}
\usepackage{amsmath,verbatim,latexsym,amssymb,indentfirst,mathrsfs,mathtools,amsthm,bbm,bm,hyperref,url,cancel,subcaption}
\usepackage[font=small,labelfont=bf,format=plain,justification=raggedright,singlelinecheck=false]{caption}
\usepackage{verbatim,indentfirst}
\usepackage[title,titletoc]{appendix}

\usepackage{graphicx}
\usepackage{epstopdf}
\newcommand{\caphead}[1]{{\bf #1}}

\renewcommand{\thesection}{\Roman{section}}
\renewcommand{\thesubsection}{\Roman{section} \Alph{subsection}}
\renewcommand{\thesubsubsection}{\Roman{section} \Alph{subsection} \arabic{subsubsection}}
\makeatletter
\def\p@subsection{}
\makeatother
\makeatletter
\def\p@subsubsection{}
\makeatother

\makeatletter
\newcommand\footnoteref[1]{\protected@xdef\@thefnmark{\ref{#1}}\@footnotemark}
\makeatother

\usepackage{soul}

\usepackage{color}
\usepackage[dvipsnames]{xcolor}
\usepackage[normalem]{ulem}

\newcommand{\KL}{{\rm KL}}

\def\id{\mathbbm{1}}   
\newcommand{\kB}{k_\mathrm{B}}  

\newcommand{\Sites}{N}  

\newcommand{\LParen}{ \bm{(} }
\newcommand{\RParen}{ \bm{)} }

\newcommand*{\Set}[1]{\left\{  #1  \right\}}

\renewcommand\th{ {\rm th} }

\begin{document}
\title{Quantifying many-body learning far from equilibrium with representation learning}
%
\author{Weishun Zhong}
\email{wszhong@mit.edu. The first two coauthors contributed equally.}
\affiliation{Physics of Living Systems, Department of Physics, Massachusetts Institute of Technology, 400 Tech Square, Cambridge, MA 02139, USA}
\author{Jacob M. Gold}
\email{jacobmg@mit.edu}
\affiliation{Department of Mathematics, Massachusetts Institute of Technology, Cambridge, Massachusetts 02139, USA}
\author{Sarah Marzen}
\email{smarzen@kecksci.claremont.edu}
\affiliation{Physics of Living Systems, Department of Physics, Massachusetts Institute of Technology, 400 Tech Square, Cambridge, MA 02139, USA}
\affiliation{W. M. Keck Science Department,
Pitzer, Scripps, and Claremont McKenna Colleges,
Claremont, CA 91711, USA}
\author{Jeremy L. England}
\email{j@englandlab.com}
\affiliation{Physics of Living Systems, Department of Physics, Massachusetts Institute of Technology, 400 Tech Square, Cambridge, MA 02139, USA}
\affiliation{GlaxoSmithKline AI/ML, 200 Cambridgepark Drive, Cambridge MA, 02140, USA}
\author{Nicole Yunger Halpern}
\email{nicoleyh@g.harvard.edu}
\affiliation{ITAMP, Harvard-Smithsonian Center for Astrophysics, Cambridge, MA 02138, USA}
\affiliation{Department of Physics, Harvard University, Cambridge, MA 02138, USA}
\affiliation{Research Laboratory of Electronics, Massachusetts Institute of Technology, Cambridge, Massachusetts 02139, USA}
\date{\today}

%
%
\begin{abstract}
Far-from-equilibrium many-body systems, from soap bubbles to suspensions to polymers, learn the drives that push them. This learning has been observed via thermodynamic properties, such as work absorption and strain. We move beyond these macroscopic properties that were first defined for equilibrium contexts: We quantify statistical mechanical learning with machine learning. Our toolkit relies on a structural parallel that we identify between far-from-equilibrium statistical mechanics and representation learning, which is undergone by neural networks that contain bottlenecks, including variational autoencoders. We train a variational autoencoder, via unsupervised learning, on configurations assumed by a many-body system during strong driving. We analyze the neural network's bottleneck to measure the many-body system's classification ability, memory capacity, discrimination ability, and novelty detection. Numerical simulations of a spin glass illustrate our technique. This toolkit exposes self-organization that eludes detection by thermodynamic measures, more reliably and more precisely identifying and quantifying learning by matter.
\end{abstract}

{\let\newpage\relax\maketitle}

%
%
%
%

Systems given many degrees of freedom
can learn and remember patterns of forces
that propel them far from equilibrium.
Such behaviors have been predicted and observed in many settings,
from charge-density waves~\cite{Coppersmith_97_Self,Povinelli_99_Noise}
to non-Brownian suspensions~\cite{Keim_11_Generic,Keim_13_Multiple,Paulsen_14_Multiple}, 
polymer networks~\cite{Majumdar_18_Mechanical}, 
soap-bubble rafts~\cite{Mukherji_19_Strength},
and macromolecules~\cite{Zhong_17_Associative}.
Such learning holds promise for engineering materials
capable of memory and computation.
This potential for applications, with experimental accessibility and ubiquity,
have earned these classical nonequilibrium many-body systems much attention recently~\cite{Keim_19_Memory}.
We measure many-body learning 
using a neural network (NN) that undergoes representation learning,
a type of machine learning.
Our toolkit detects and quantifies many-body learning
more thoroughly and precisely
than thermodynamic measures used to date.

One of the 
best-characterized instances of learning by driven matter
involves a  spin glass.
The spins are classical and interact randomly.
Consider applying fields from a set $\{ \vec{A}, \vec{B}, \vec{C} \}$,
which we call a \emph{drive}.
As the driving proceeds, the spins flip, absorbing work.
In a certain parameter regime,
the absorbed power shrinks adaptively:
The spins migrate toward a corner of configuration space 
where their configuration withstands the drive's insults.
Consider then imposing fields absent from the original drive.
Subsequent spin flips will absorb more work
than if the field belonged to the original drive.
Insofar as a simple, low-dimensional property of the material 
can be used to discriminate between 
drive inputs that fit a pattern and drive inputs that do not, 
we say that the material has learned the drive.

Learning behavior has been quantified with 
properties commonplace in thermodynamics.
Examples include work, magnetization, and strain.
This thermodynamic characterization has provided insights
but suffers from two shortcomings.
First, the types of thermodynamic properties vary from system to system.
For example, work absorption characterizes the spin glass's learning;
strain characterizes polymer networks'.
A more general approach would facilitate comparisons and standardize analyses.
Second, thermodynamic properties are useful 
for characterizing macroscopic equilibrium states.
But such properties are not necessarily the best
for describing the far-from-equilibrium systems that learn.

Over the past several years, machine learning has revolutionized 
the quantification of learning~\cite{Nielsen_15_Neural,Goodfellow_16_Deep}.
Machine learning calls for application
to the learning of drive patterns by many-body systems.

\begin{figure}[hbt]
\centering
\includegraphics[width=.25\textwidth, clip=true]{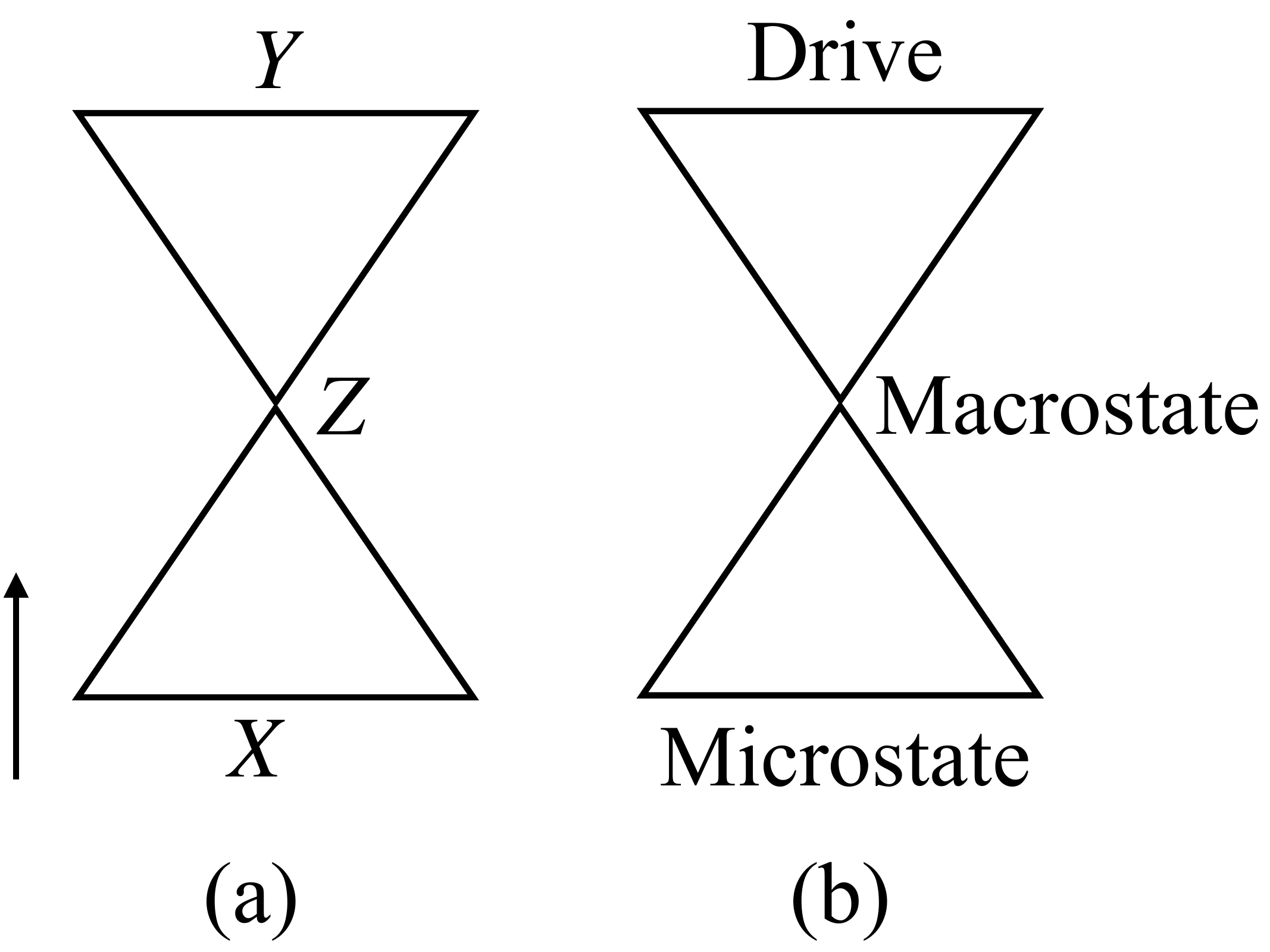}
\caption{\caphead{Parallel between two structures:}
(a) Structure of a bottleneck neural network, which performs representation learning.
(b) Structure of a far-from-equilibrium-statistical-mechanics problem.}
\label{fig_VAE_SM_Parallel}
\end{figure}

Parallels between statistical mechanics and 
certain machine-learning components 
have been known for decades~\cite{Engel_01_Statistical,Nielsen_15_Neural}.
For example, Boltzmann machines resemble
particles exchanging heat with thermal baths.
Parallels between \emph{representation learning} and statistical mechanics
have enjoyed less attention
(though one parallel was proposed in~\cite{Alemi_18_TherML}).
Figure~\ref{fig_VAE_SM_Parallel}(a) illustrates 
representation learning~\cite{Bengio_12_Representation}:
A high-dimensional variable $X$ is inputted into a NN.
The NN compresses relevant information 
into a low-dimensional variable $Z$.
The NN then decompresses $Z$ into a prediction $\hat{Y}$
of a high-dimensional variable $Y$.
If $Y = X$, the NN is an autoencoder,
mimicking the identity function.
The latent variable $Z$ acts as a bottleneck.
The bottleneck's size controls a tradeoff
between the memory consumed and the prediction's accuracy.
We call the NNs that perform representation learning
\emph{bottleneck NNs}.

Representation learning, we argue, shares its structure with
problems in which a strong drive forces a many-body system
[Fig.~\ref{fig_VAE_SM_Parallel}(b)].
The system's microstate, like $X$, occupies a high-dimensional space.
A macrostate synopsizes the microstate in a few numbers, 
such as particle number and magnetization.
This synopsis parallels $Z$.
If the system has learned the drive, the macrostate encodes the drive.
One may reconstruct the drive from the macrostate,
as a bottleneck NN reconstructs $Y$ from $Z$.

Applying this analogy, we use representation learning 
to measure how effectively a far-from-equilibrium many-body system
learns a drive.
We illustrate with numerical simulations of the spin glass, 
whose learning has been characterized with work absorption~\cite{Gold_19_Self}.
However, our methods generalize to other platforms.
Our measurement scheme offers three advantages:
\begin{enumerate}
   \item 
   Bottleneck NNs register learning behaviors 
   more thoroughly, reliably, and precisely
   than work absorption.

   \item 
   Our framework applies to a wide class of
   strongly driven many-body systems.
   The framework does not rely on 
   strain, work absorption, susceptibility, etc.
   Hence our toolkit can characterize spins, suspensions, polymers, etc.

   \item 
   Our approach unites a machine-learning sense of learning
   with the statistical mechanical sense.
   This union is conceptually satisfying.
   
\end{enumerate}
We apply representation learning to measure 
 classification, memory capacity, discrimination, and novelty detection.
Our techniques can be extended to other facets of learning,
such as prediction and the decomposition of a drive into constituents.

Most of our measurement schemes have the following structure:
The many-body system is trained with 
a drive (e.g., fields $\vec{A}$, $\vec{B}$, and $\vec{C}$).
Then, the system is tested (e.g., with a field $\vec{D}$).
Training and testing are repeated in many trials.
Configurations realized 
are used to train a bottleneck NN.
In some cases, the NN then receives data from
the statistical mechanical testing.
Finally, we analyze the NN's latent space and/or predictions.

The rest of this paper is organized as follows.
Section~\ref{sec_Setup} introduces the bottleneck NN that we use
and the spin-glass example.
In Sec.~\ref{sec_Results},
we prescribe how to quantify, using representation learning, 
the learning of a drive by a many-body system.
Section~\ref{sec_Discussion} closes with a discussion:
We decode our NN's latent space in terms of thermodynamic variables,
argue for our techniques' feasibility,
and detail opportunities engendered by this study.

%
%
%
%
%
\section{Setup: Representation-learning model
and spin-glass example}
\label{sec_Setup}

This section introduces two toolkits applied in Sec.~\ref{sec_Results}:
(i) Section~\ref{sec_Intro_NNs} details the bottleneck NN we use.
(ii) Section~\ref{sec_Spin_Glass} details the spin glass
with which we illustrate statistical mechanical learners.

\subsection{Representation-learning model}
\label{sec_Intro_NNs}


This section overviews our architecture;
details appear in App.~\ref{sec_NN_Details}.
This paper's introduction identifies a parallel between
thermodynamic problems and bottleneck NNs (Fig.~\ref{fig_VAE_SM_Parallel}).
In the thermodynamic problem, $Y \neq X$ represents the drive.
We could design a bottleneck NN that predicts drives from configurations $X$.
But the NN would need to undergo supervised learning,
if built according to today's standards.
During supervised learning, the NN would receive tuples 
(configuration, label of drive that generated the configuration).
Receiving drive labels would give the NN
information not directly accessible to the many-body system.
The NN's predictive success would not necessarily reflect
only learning by the many-body system.
Hence we design a bottleneck NN that performs unsupervised learning,
receiving just configurations.

This NN is a \emph{variational autoencoder} (VAE)~\cite{Kingma_13_Auto,JR_14_Stochastic,Doersch_16_Tutorial}.
A VAE is a generative model:
It receives samples $x$ from a distribution over
the possible values of $X$,
learns about the distribution, and generates new samples.
The NN approximates the distribution,
using Bayesian variational inference (App.~\ref{sec_NN_Details}).
The parameters are optimized during training facilitated by backpropagation.

Our VAE has five fully connected hidden layers,
with neuron numbers 200-200-(number of $Z$ neurons)-200-200.
We usually restrict $Z$ to 2-4 neurons.
This choice facilitates the visualization of the latent space
and suffices to quantify our spin glass's learning.
Growing the number of degrees of freedom,
and the number of drives, may require more dimensions.
But our study suggests that the number of dimensions needed
$\ll$ the system size.

\begin{figure}[hbt]
\centering
\includegraphics[width=.45\textwidth, clip=true]{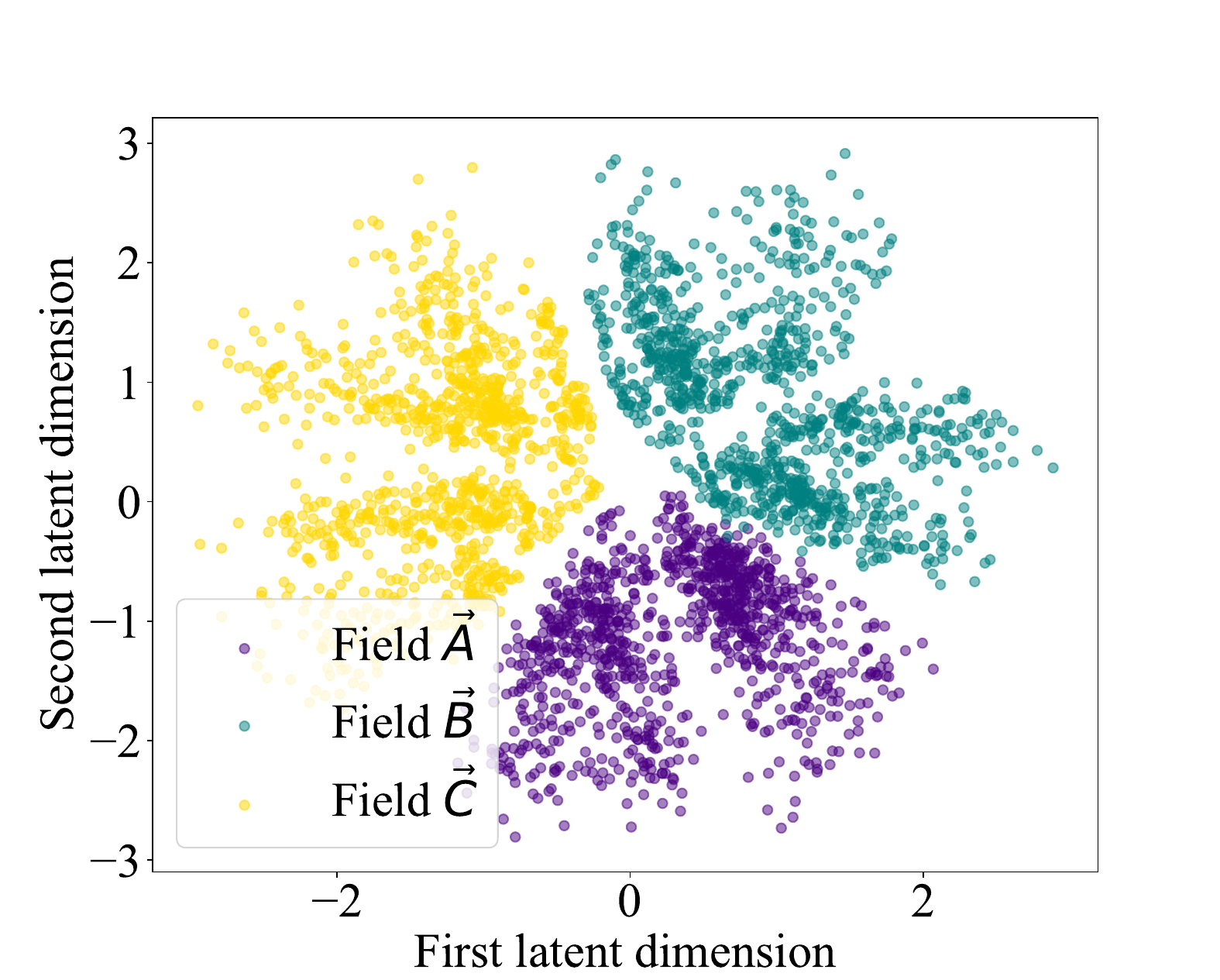}
\caption{\caphead{Visualization of latent space:}
The latent space $Z$ consists of two neurons, $Z_1$ and $Z_2$.
A variational autoencoder (VAE) formed this latent space
while training on configurations assumed by a 256-spin glass
during repeated exposure to three fields, $A$, $B$, and $C$. 
The VAE mapped each configuration to a dot in latent-space.
We color each dot in accordance with
the field that produced the configuration.
Same-color dots cluster together:
The VAE identified which configurations resulted from the same field.}
\label{fig_Latent_Space}
\end{figure}

The latent space is visualized in Fig.~\ref{fig_Latent_Space}.
Each neuron corresponds to one axis
and represents a continuous-valued real number.
The VAE maps each inputted configuration
to one latent-space dot. 
Close-together dots correspond to
configurations produced by the same field,
if the spin glass and NN learn well.
We illustrate this clustering in Fig.~\ref{fig_Latent_Space}
by coloring each dot according to the drive that produced it.

%
%
\subsection{Spin glass}
\label{sec_Spin_Glass}


A spin glass exemplifies the statistical mechanical learner.
We introduce the spins, Hamiltonian, and heat exchange below.
We model the time evolution, define work and heat,
and describe the initialization procedure.
Finally, we describe a parameter regime
in which the spin glass learns effectively.
Appendix~\ref{sec_Not_Enslaved_Or_Frozen}
distinguishes robust learning from superficially similar behaviors.

We adopt the model in~\cite{Gold_19_Self}.
Simulations are of $\Sites = 256$ classical spins.
The $j^\th$ spin occupies one of two possible states:
$s_j = \pm 1$.

The spins couple together and experience an external magnetic field.
Spin $j$ evolves under a Hamiltonian
\begin{align}
   \label{eq_Hamiltonian_j}
   H_j(t)
   =  \sum_{k \neq j}  J_{jk}  s_j  s_k
   +  A_j(t) s_j ,
\end{align}
and the spin glass evolves under
\begin{align}
   \label{eq_Hamiltonian}
   H(t)  
   =  \frac{1}{2} \sum_{j = 1}^\Sites  H_j(t) 
\end{align}
at time $t$.
We call the first term in Eq.~\eqref{eq_Hamiltonian_j} the \emph{interaction energy} 
and the second term the \emph{field energy}.
The couplings $J_{j k}  =  J_{kj}$ are defined in terms of 
an Erd\"{o}s-R\'enyi random network:
Nodes $j$ and $k$ have some probability $p$ 
of sharing an edge, for all $j$ and $k \neq j$.
We identify nodes with spins and identify edges with couplings.
Each spin couples to eight other spins, on average.
The nonzero couplings $J_{j k}$ are selected according to 
a normal distribution of standard deviation 1.

\begin{figure}[hbt]
\centering
\includegraphics[width=.5\textwidth, clip=true]{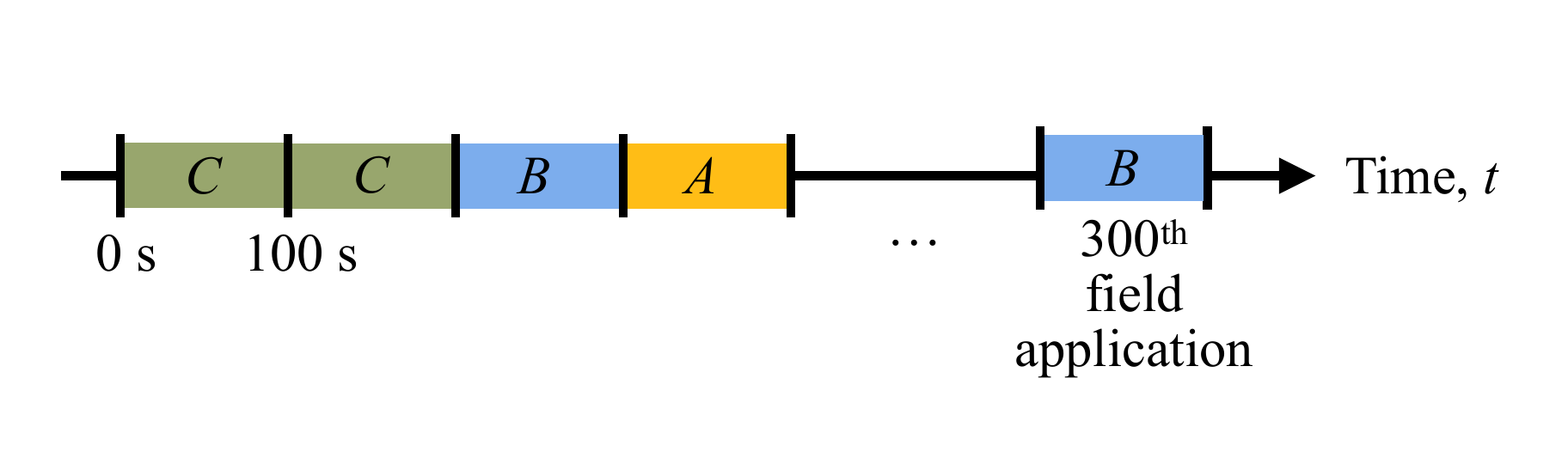}
\caption{\caphead{Driving protocol:} 
The drive consists of the set $\{A, B, C\}$ of fields.
A field is selected randomly from the drive and applied for 100 s,
and then this process is repeated.}
\label{fig_Drive_Protocol}
\end{figure}

The $A_j(t)$ in Eq.~\eqref{eq_Hamiltonian_j} 
is defined as follows.
At time $t$, the spin glass experiences a field $\{ A_j(t) \}_j$.
$A_j(t)$ represents the magnitude and sign of the field at spin $j$.
All fields point along the same direction (conventionally labeled as the $z$-axis), 
so we simplify the vector notation $\vec{A}_j$ to $A_j$.
Elsewhere in the text, we simplify $\{ A_j (t) \}_j$
to the capital Latin letter $A$ (or $B$, or $C$, etc.). 
Each $A_j(t)$ is selected according to 
a normal distribution of standard deviation 3.
The field changes every 100 seconds. 
To train the spin glass, we construct a drive
by forming several random fields $\{ A_j \}_j$.
We randomly select a field from the set, then apply the field.
We repeat these two steps 299 times, unless otherwise noted
(Fig.~\ref{fig_Drive_Protocol}).

The spin glass exchanges heat with
a bath at a temperature $T = 1 / \beta$.
We set Boltzmann's constant to one: $\kB = 1$.
Energies are measured in units of Kelvins (K).
To flip, a spin must overcome an energy barrier of height $B$.
Spin $j$ tends to flip at a rate 
\begin{align}
   \label{eq_Flip_Rate}
   \omega_j 
   =  e^{\beta [ H_j(t) - B]}
   / (1 \text{ second})  \, .
\end{align}
Equation~\eqref{eq_Flip_Rate} has the form of Arrhenius's law
and obeys detailed balance.
Each spin flips once every $10^7$ s, on average.
We model the evolution with discrete 100-s time intervals,
using the Gillespie algorithm.

The spins absorb work when the field changes,
and they dissipate heat while flipping,
as we now detail.
Consider changing the field from $\{ A_j (t) \}$ to $\{ A'_j(t) \}$. 
The change in the spin glass's energy equals
the work absorbed by the spin glass:
\begin{align}
   W  :=  \sum_{j = 1}^\Sites  
   \left[  A'_j(t)  -  A_j(t)  \right]  s_j .
\end{align}
To define heat, we suppose that spin $k$ flips at time $t$:
$s_k  \mapsto  s'_k  =  - s_k$. 
The spin glass dissipates an amount $Q$ of heat equal to 
the negative of the change in the spin glass's energy:
\begin{align}
   Q  
   & :=  -  \frac{1}{2}  \sum_{j \neq k}  \left[
   J_{jk}  s_j  (s'_k - s_k)
   +  A_k(t)  (s'_k  -  s_k)  \right]  \\
   & =  \sum_{j \neq k}  J_{j k}  s_j  s_k
   +  2 h_k (\alpha_t)  s_k .
\end{align}
Our discussion is cast in terms of the absorbed power,
$W / ( \text{100 s} )$. 

The spin glass is initialized in a uniformly random configuration $C$.
Then, the spins relax in the absence of any field for 100,000 seconds.
The spin glass navigates to near a local energy minimum.
If a protocol is repeated in multiple trials, 
all the trials begin with the same $C$.

In a certain parameter regime, the spin glass learns its drive effectively,
even according to the absorbed power~\cite{Gold_19_Self}.
Consider training the spin glass on a drive $\{ A, B, C \}$.
The spin glass absorbs much work initially.
If the spin glass learns the drive, the absorbed power declines.
If a dissimilar field $D$ is then applied, the absorbed power spikes.
The spin glass learns effectively when
$\beta = 3$ K$^{-1}$ and $B = 4.5$ K~\cite{Gold_19_Self}.
These parameters define a Goldilocks regime:
The temperature is high enough,
and the barriers are low enough,
that the spin glass can explore phase space.
But $T$ is low enough, and the barriers are high enough,
that the spin glass is not hopelessly peripatetic.

\section{How to detect and quantify \\ a many-body system's \\ learning of a drive,
using representation learning}
\label{sec_Results}

This section shows how to quantify four facets of learning.
Section~\ref{sec_Classify} concerns
the many-body system's ability to classify drives; 
Sec.~\ref{sec_Capacity}, memory capacity;
Sec.~\ref{sec_Discriminate}, discrimination of similar fields;
and Sec.~\ref{sec_ROC}, novelty detection.
At the end of each section, we synopsize 
the technique introduced in boldface.
These four techniques illustrate how representation learning 
can be applied to quantify features of learning.
Other features may be quantified along similar lines.
Code used can be found at the online repository~\cite{Github_repo}.

\subsection{Classification: Which drive is this?}
\label{sec_Classify}




A system classifies a drive by identifying the drive as 
one of many possibilities.
A VAE, we find, reflects more of a spin glass's classification ability
than absorbed power does.

We illustrate with the spin glass.
We generated random fields $A$, $B$, $C$, $D$, and $E$.
From 4 of the fields, we formed the drive $\mathcal{D}_1  :=  \{A, B, C, D\}$.
On the drive, we trained the spin glass in each of 1,000 trials.
In each of 1,000 other trials, we trained a fresh spin glass on
a drive $\mathcal{D}_2  :=  \{A, B, C, E\}$.
We repeated this process for each of the 5 possible 4-field drives.
Ninety percent of the trials were randomly selected for training our NN.
The rest were used for testing.

We used the VAE to gauge the spin glass's classification of the drives:
We identified the configurations 
occupied by the spin glass at a fixed time $t$ in the training trials.
On these configurations, we trained the VAE.
The VAE populated the latent space with dots
(as in Fig.~\ref{fig_Latent_Space})
whose density formed a probability distribution.

\begin{figure}[hbt]
\centering
\includegraphics[width=.47\textwidth, clip=true]{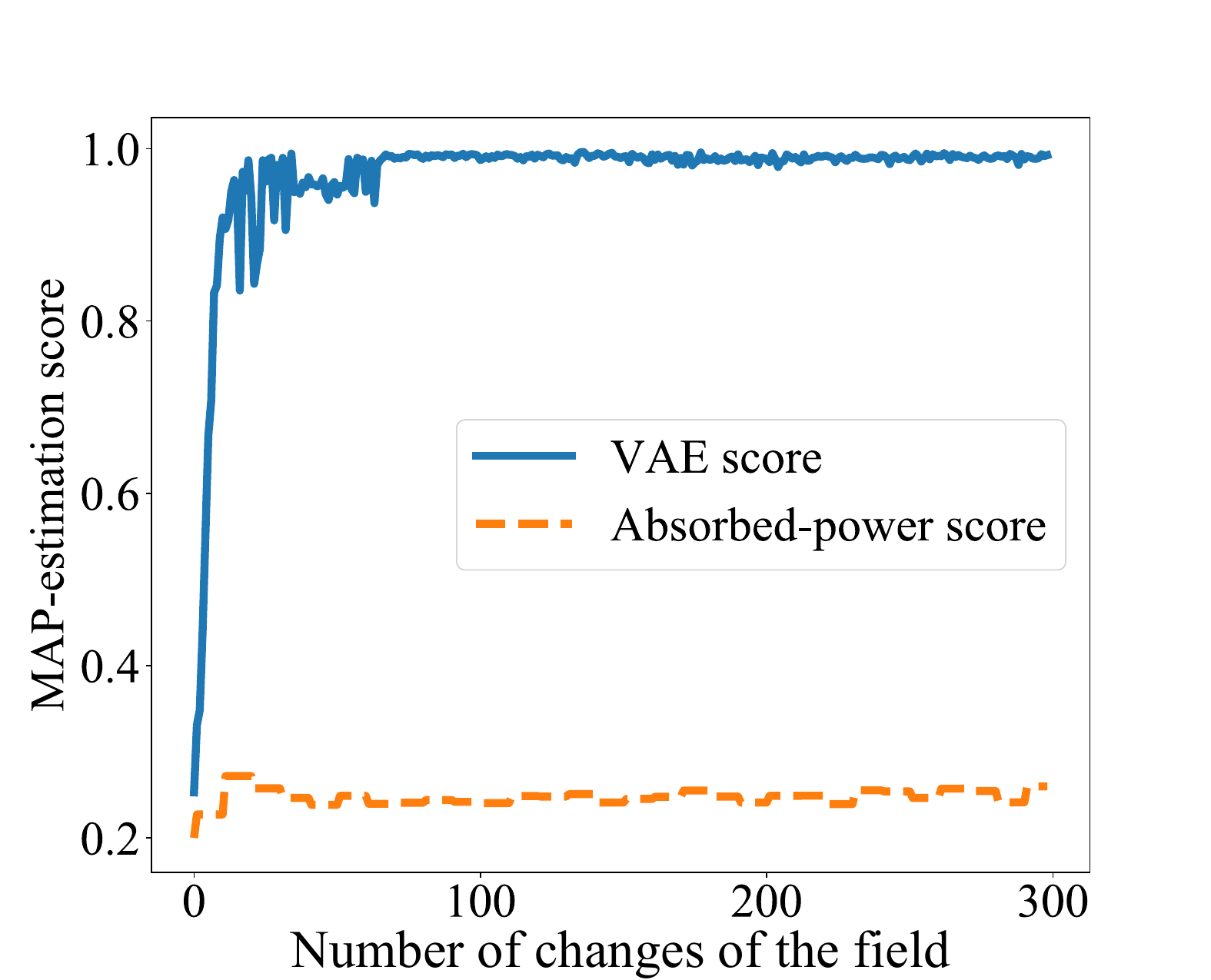}
\caption{\caphead{Quantification of a many-body system's classification ability:}
A spin glass classified a drive as one of five possibilities.
The system's classification ability was defined as
the score of the maximum \emph{a posteriori} (MAP) estimation
performed with a variational autoencoder (VAE) (blue, upper curve).
We compare with the score of MAP estimation performed with
absorbed power (orange, lower curve).
The VAE score rises to near the maximal value, 1.
The thermodynamic score remains slightly above 
the random-guessing score, $0.20$.
Hence the VAE detects more of the spin glass's classification ability.
}
\label{fig_Classification}
\end{figure}

We then showed the VAE a time-$t$ configuration
from a test trial.
The VAE compressed the configuration into a latent-space point.
We calculated which drive most likely, 
according to the probability density,
generated the latent-space point.
The calculation was \emph{maximum} a posteriori \emph{estimation} 
(MAP estimation) (see~\cite{Bishop_06_Pattern} and App.~\ref{app_MAP}).
Here, the MAP estimation is equivalent to maximum-likelihood estimation.
We performed this testing and estimation for each trial in the test data.
The fraction of trials in which MAP estimation succeeded
forms the \emph{score}.
We scored the classification at each of many times $t$.
The score is plotted against $t$ in Fig.~\ref{fig_Classification},
as the blue, upper curve.

The absorbed power reflects 
the spin glass's classification of the drives as follows.
For each drive $\mathcal{D}$ and each time $t$,
we histogrammed the power absorbed
while $\mathcal{D}$ was applied at $t$
in a VAE-training trial.
Then, we took a trial from the test set
and identified the power $\mathcal{P}$ absorbed at $t$.
We inferred which drive most likely, 
according to the histograms, produced $\mathcal{P}$. 
The guess's score appears as the orange, lower curve
in Fig.~\ref{fig_Classification}.

The score maximizes at 1.00 if the drive is always guessed accurately.
The score is lower-bounded by $1 / (\text{number of drives}) = 0.20$,
which results from random guessing.
In Fig.~\ref{fig_Classification}, each score grows 
over a time scale of tens of field switches.
The absorbed-power score begins at 0.20\footnote{
\label{foot_Why_Not_0.2}
The VAE's score begins close to 0.20. The slight distance from 0.20, 
we surmise, comes from stochasticity of three types:
the spin glass's initial configuration, the MAP estimation,
and stochastic gradient descent. 
Stochasticity of only the first two types affects the absorbed power's score.
}
and comes to fluctuate around 0.25.
The VAE's score comes to fluctuate slightly below 1.00.
Hence the VAE reflects more of the spin glass's classification ability
than the absorbed power does.

\textbf{A many-body system's ability to classify drives is quantified with 
the score of MAP estimates calculated from a VAE's latent space.}

\subsection{Memory capacity: How many drives can be remembered?}
\label{sec_Capacity}


How many fields can a many-body system remember?
A VAE, we find, registers a greater capacity 
than absorbed power registers.\footnote{
We use the term ``memory capacity'' in the physical sense of~\cite{Keim_19_Memory}.
A more specific, technical definition of ``memory capacity''
is used in reservoir computing~\cite{Jaeger_02_Short}.
}
Hence the VAE reflects statistical mechanical learning,
at high field numbers,
that the absorbed power does not.

We illustrated by constructing 50 random fields.
We selected 40 to form a drive $\mathcal{D}_1$,
selected 40 to form a drive $\mathcal{D}_2$,
and repeated until forming 5 drives.
We trained the spin glass on $\mathcal{D}_j$
in each of 1,000 trials, for each of $j = 1, 2, \ldots 5$.
Ninety percent of the trials were designated as VAE-training trials;
and 10\%, as VAE-testing trials.

The choice of 50 fields is explained in App.~\ref{app_Capacity_Work}:
Fifty fields exceed the spin-glass capacity registered by the absorbed power.
We aim to show that 50 fields do not exceed the capacity
registered by the VAE:
The VAE identifies spin-glass learning missed by the absorbed power.

\begin{figure}[hbt]
\centering
\includegraphics[width=.45\textwidth, clip=true]{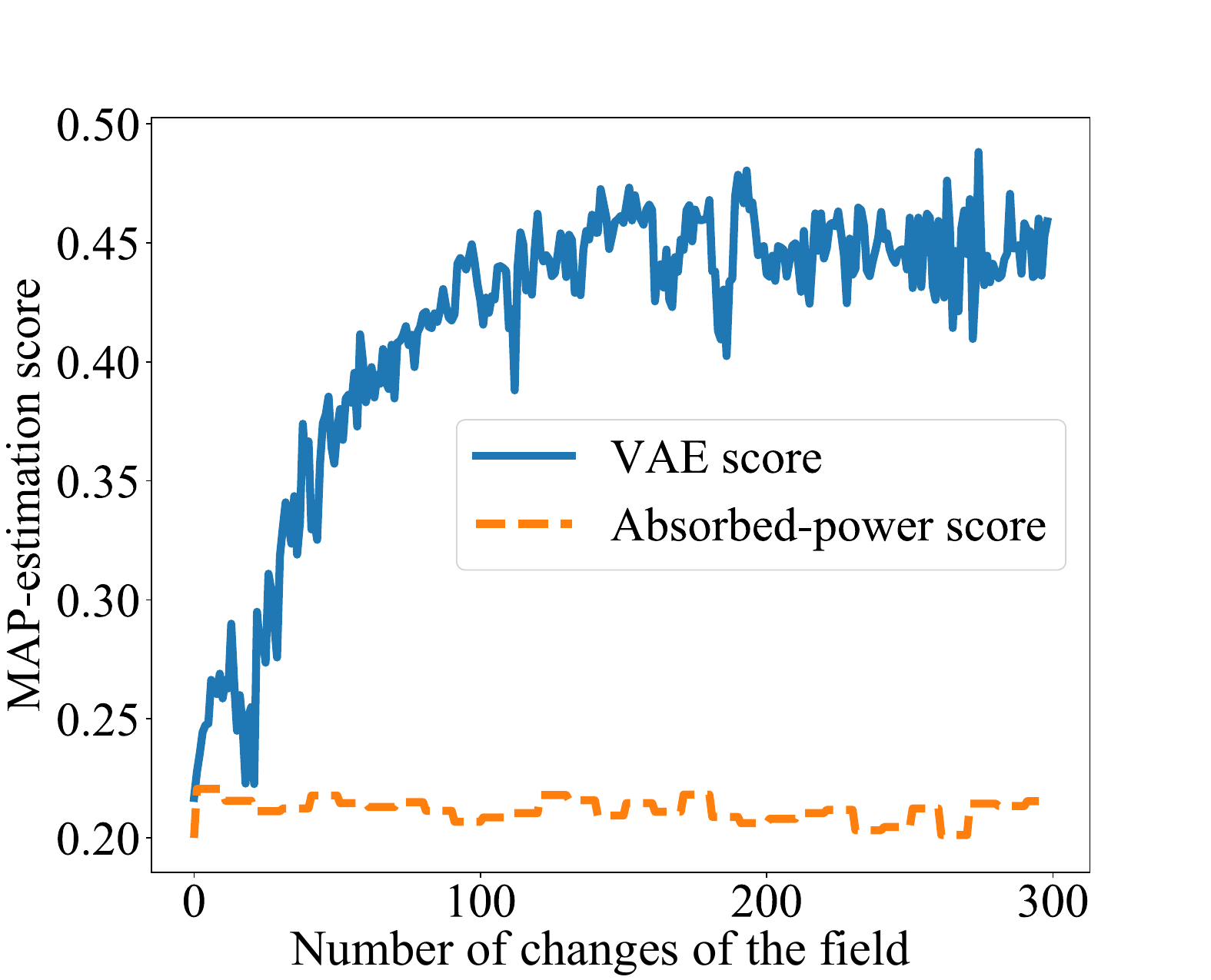}
\caption{\caphead{Quantification of memory capacity:}
A spin glass was trained on one of five drives in each of many trials.
Each drive was formed from 40 fields selected from 50 fields.
We quantified the spin glass's ability to classify the drives
with the score of maximum \emph{a priori} (MAP) estimation
performed with a variational autoencoder (upper, blue line).
The score of MAP estimation performed with absorbed power
is shown for comparison (lower, orange line).}
\label{fig_Capacity}
\end{figure}

We used representation learning to quantify the spin glass's capacity as follows.
For a fixed time $t$, we collected 
the configurations occupied by the spin glass at $t$
in the VAE-training trials.
On these configurations, the VAE performed unsupervised learning.
The VAE populated its latent space with dots
that formed five clusters.
Then, we fed the VAE the configuration occupied at $t$ during a test trial.
The VAE formed a new dot in latent space.
We MAP-estimated the drive that, according to the VAE,
most likely generated the dot (Sec.~\ref{sec_Classify}).
The fraction of test trials in which the VAE guessed correctly
constitutes the VAE's score.
The score is plotted against $t$ in Fig.~\ref{fig_Capacity},
as the blue, upper curve.

The VAE's score is compared with the absorbed power's score,
calculated as follows.
For a fixed time $t$, we identified the power absorbed at $t$
in each VAE-testing trial.
We histogrammed the power absorbed
when $\mathcal{D}_j$ was applied at $t$,
for each $j = 1, 2, \ldots, 5$.
We then identified the power absorbed at $t$ in a test trial.
Comparing with the histograms, we inferred which drive
was most likely being applied.
We repeated this inference with each other test trial.
In which fraction of the trials did the absorbed power identify the drive correctly?
This number forms the absorbed power's score.
The score is plotted as the lower, orange curve in Fig.~\ref{fig_Capacity}.

The higher the score, the greater the memory capacity
attributed to the spin glass.
The absorbed power identifies the drive 
in approximately $20\%$ of the trials,
as would random guessing.
The score remains approximately constant,
because the number of fields exceeds the spin-glass capacity
measured by the absorbed power.
In contrast, the VAE's score grows over
$\approx 150$ changes of the field,
then plateaus at $\approx 0.450$.
The VAE points to the wrong drive most of the time
but succeeds significantly more often than the absorbed power.
Hence representation learning uncovers
more of the spin glass's memory capacity
than absorbed power measure does.

\textbf{A many-body system's memory capacity is quantified with
the greatest number of fields in any drive 
on which MAP estimation, based on a VAE's latent space,
scores better than random guessing.}


%
%
%
\subsection{Discrimination: How new is this field?}
\label{sec_Discriminate}


Suppose that a many-body system learns fields $A$ and $B$,
then encounters a field that interpolates between them.
Can the system recognize that 
the new field contains familiar constituents?
Can the system discern how much $A$ contributes
and how much $B$ contributes?
The answers characterize the system's discrimination ability,
which we quantify with a MAP-estimation score (Sec.~\ref{sec_Classify}).
Estimates formed from a VAE's latent space
reflect more of the system's discriminatory ability
than do estimates formed from absorbed power.

We illustrate with the spin glass, forming a drive $\{A, B, C\}$.
In each of 300 time intervals,
a field was selected randomly from the drive and applied.
The spin glass was then tested with
a linear combination $D_w = w A + (1 - w) B$.
The weight $w$ varied from 0 to 1, in steps of $1/6$, across trials.

We measured the spin glass's discrimination using the VAE as follows.
The final configuration assumed by the spin glass 
in each test trial was collected.
The configurations were split into VAE-training data
and VAE-testing data.
On the configurations generated by $D_w$ in the VAE-training data,
the VAE was trained.
Then, the VAE received a configuration generated by $D_w$
in a VAE-testing trial.
The VAE mapped the configuration to a latent-space point.
We inferred which field most likely generated that point,
using MAP estimation on the latent space.
We tested the VAE many times, 
then calculated the fraction of MAP estimates that were correct,
the VAE's score.

Similarly, we measured the spin glass's discrimination
using the absorbed power.
For each trial in the VAE-training data,
we calculated the power $\mathcal{P}$ absorbed by the spin glass
after the application of $D_w$.
We histogrammed $\mathcal{P}$, inferring the probability that,
if shown $D_w$ for a given $w$,
the spin glass will absorb an amount $\mathcal{P}$ of power.
Then, we calculated the power absorbed
during a VAE-testing trial.
We inferred which field most likely generated that point,
using MAP estimation on the latent space.
Repeating MAP estimation with all the VAE-testing trials,
we calculated the absorbed power's score.

The VAE's score equals about double the absorbed power's score,
for latent spaces of dimensionality 2 to 20.
The VAE scores between 0.448 and 0.5009, 
whereas the absorbed power scores 0.2381.
Hence the representation-learning model picks up on 
more of the spin glass's discriminatory ability
than the absorbed power does.

\textbf{A many-body system's ability to discriminate
combinations of familiar fields
is quantified with the score of MAP estimates
formed from a VAE's latent space.}

\subsection{Novelty detection: Has this drive been encountered before?}
\label{sec_ROC}


\begin{figure}[hbt]
\centering
\includegraphics[width=.45\textwidth, clip=true]{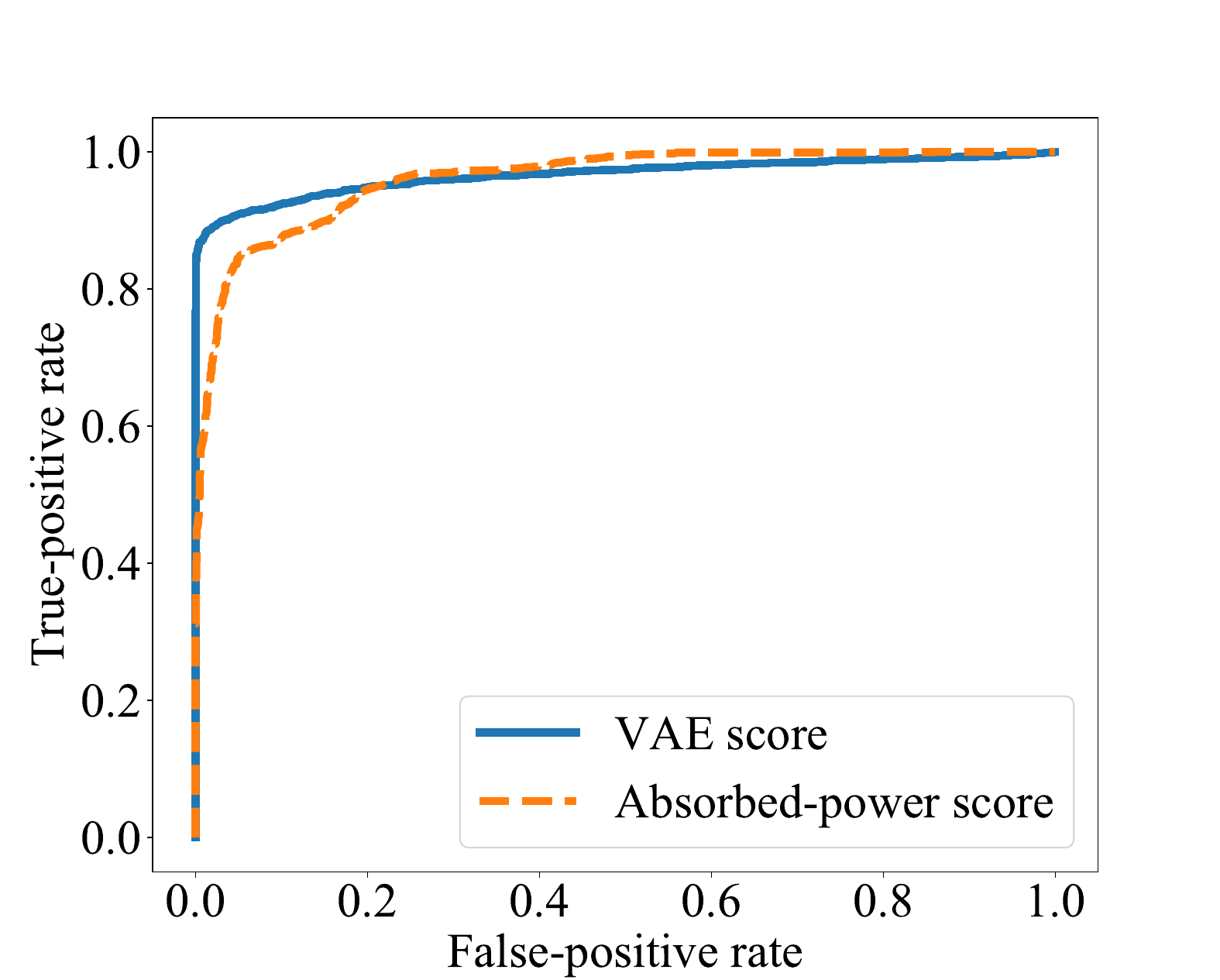}
\caption{\caphead{Receiver-operating-characteristic (ROC) curve:}
The spin glass was trained with three drives,
then tested with a familiar drive or with a novel drive.
From a response of the system's, an ROC curve can be defined.
The blue, solid curve is defined in terms of a variational autoencoder;
and the orange, dashed curve is defined in terms of absorbed power.
}
\label{fig_ROC}
\end{figure}

At the start of the introduction, we described
how absorbed power has been used to identify novelty detection.
A system detects novelty when labeling a stimulus as familiar or unfamiliar.
The stimulus produces a response that exceeds a threshold or lies below.
If the stimulus exceeds the threshold, 
an observer should guess that the stimulus is novel.
Otherwise, the observer should guess that the stimulus is familiar.

The observer can err in two ways:
One commits a \emph{false positive} by believing a familiar drive to be novel.
One commits a \emph{false negative} by believing a novel drive to be familiar.
The errors trade off:
Raising the threshold lowers the probability
$p( \text{pos.} | \text{neg.} )$,
suppressing false positives at the cost of false negatives.
Lowering the threshold lowers the probability 
$p( \text{neg.} | \text{pos.} )$,
suppressing false negatives at the cost of false positives.

The \emph{receiver-operating-characteristic} (ROC) curve
depicts the tradeoff's steepness 
(see~\cite{Brown_06_Receiver} and Fig.~\ref{fig_ROC}).
Each point on the curve corresponds to one threshold value.
The false-positive rate $p( \text{pos.} | \text{neg.} )$ 
runs along-the $x$-axis; and the true-positive rate, 
$p( \text{pos.} | \text{pos.} )$, along the $y$-axis.
The greater the area under the ROC curve,
the more sensitively the response reflects accurate novelty detection.

We measure a many-body system's 
novelty-detection ability using an ROC curve.
Let us illustrate with the spin glass.
We constructed two random drives,
$\{A, B, C \}$ and $\{D, E, F \}$.
We trained the spin glass on $\{A, B, C \}$.
In each of 3,000 trials, we then tested the spin glass with
$A$, $B$, or $C$.
In each of 3,000 other trials, we tested with $D$, $E$, or $F$.
We defined one response in terms of a VAE, as detailed below;
measured the absorbed power, a thermodynamic response;
and, from each response, drew an ROC curve
(Fig.~\ref{fig_ROC}).
The curves show that representation learning offers 
greater sensitivity to the spin glass's novelty detection.

We defined the representation-learning response in terms of a VAE as follows.
We trained the VAE on the configurations 
assumed by the spin glass during its training.
The VAE populated latent space with three clumps of dots.
We modeled the clumps with a hard mixture 
$p_{ABC} (z_1,  z_2)$ of three Gaussians.\footnote{
A mixture is hard if it models each point as
belonging to only one Gaussian.}
We then fed the VAE the configuration that resulted from testing the spin glass.
The VAE mapped the configuration to a latent-space point 
$(z_1^{\rm test},  z_2^{\rm test})$.
We calculated the probability 
$p_{ABC} (z_1^{\rm test},  z_2^{\rm test}) \, dz_1 dz_2$ 
that the $ABC$ distribution produced the new point.
This protocol led to the blue, solid curve in Fig.~\ref{fig_ROC}.

We defined a thermodynamic ROC curve in terms of absorbed power.
Consider the trials in which the spin glass is tested with field $A$.
We histogrammed the power absorbed by the spin glass 
at the end of the testing.
We form another histogram from the $B$-test trials;
and a third histogram, from the $C$-test trials.
To these histograms was compared
the power $\mathcal{P}$ that the spin glass absorbed
during a test with an arbitrary field.
We inferred the likelihood that $\mathcal{P}$ resulted from
a familiar field.
The results form the orange, dashed curve in Fig.~\ref{fig_ROC}.

The two ROC curves enclose regions of approximately the same area:
The VAE curve encloses an area-0.9633 region;
and the thermodynamic curve, an area-0.9601 region.
On average across all thresholds, therefore,
the responses register novelty detection approximately equally.
Yet the responses excel in different regimes:
The VAE achieves greater true-positive rates at low false-positive rates,
and the absorbed power achieves greater true-positive rates 
at high false-positive rates.
This two-regime behavior persisted across batches of trials,
though the enclosed areas fluctuated a little.
Hence anyone paranoid about avoiding false positives
should measure a many-body system's novelty detection with a VAE,
while those more relaxed might prefer the absorbed power.

Why should the VAE excel at low false-positive rates?
Because of the VAE's skill at generalizing, we conjecture.
Upon training on cat pictures, a VAE generalizes from the instances.
Shown a new cat, the VAE recognizes its catness.
Perturbing the input a little perturbs the VAE's response little.
Hence changing the magnetic field a little,
which changes the spin-glass configuration little,
should change latent space little,
obscuring the many-body system's novelty detection.
This obscuring disappears when the magnetic field changes substantially.

\textbf{A many-body system's novelty-detection ability is quantified with
an ROC curve formed from a VAE's latent space
or a thermodynamic response, depending on the false-positive threshold.}

\section{Discussion}
\label{sec_Discussion}

We have detected and quantified a many-body system's learning of its drive,
using representation learning,
with greater sensitivity than absorbed power affords.
The scheme relies on a parallel that we identified
between statistical mechanical problems and VAEs.
Uniting statistical mechanical learning with machine learning,
the definition is conceptually satisfying.
The definition also has wide applicability,
not depending on whether 
the system exhibits magnetization or strain or another thermodynamic response.
Furthermore, our representation-learning toolkit signals many-body learning
more sensitively than does
the seemingly best-suited thermodynamic tool.

The rest of this section is organized as follows.
In Sec.~\ref{sec_Decode_Latent}, 
we decode latent space in terms of thermodynamic variables.
In Sec.~\ref{sec_Feasibility}, we argue for the feasibility of applying our toolkit.
In Sec.~\ref{sec_Opportunities}, 
we discuss problems that our toolkit can illuminate.
We also motivate the development of new representation-learning tools.

\subsection{Decoding latent space}
\label{sec_Decode_Latent}

Thermodynamicists parameterize macrostates with
volume, energy, magnetization, etc.
Thermodynamic macrostates parallel latent space,
as illustrated in Fig.~\ref{fig_VAE_SM_Parallel}.
What variables parameterize the VAE's latent space?
Latent space could suggest definitions of new thermodynamic variables,
or hidden relationships amongst known thermodynamic variables.
We begin decoding latent space in terms of thermodynamic quantities,
leaving the full decoding for future research.

We illustrate with part of the spin-glass protocol 
in Sec.~\ref{sec_Classify}:
Train the spin glass with a drive $\{ A, B, C \}$ in each of many trials.
On the end-of-trial configurations, train the VAE.

\begin{figure}[h]
\centering
\begin{subfigure}{0.4\textwidth}
\centering
\includegraphics[width=1\textwidth]{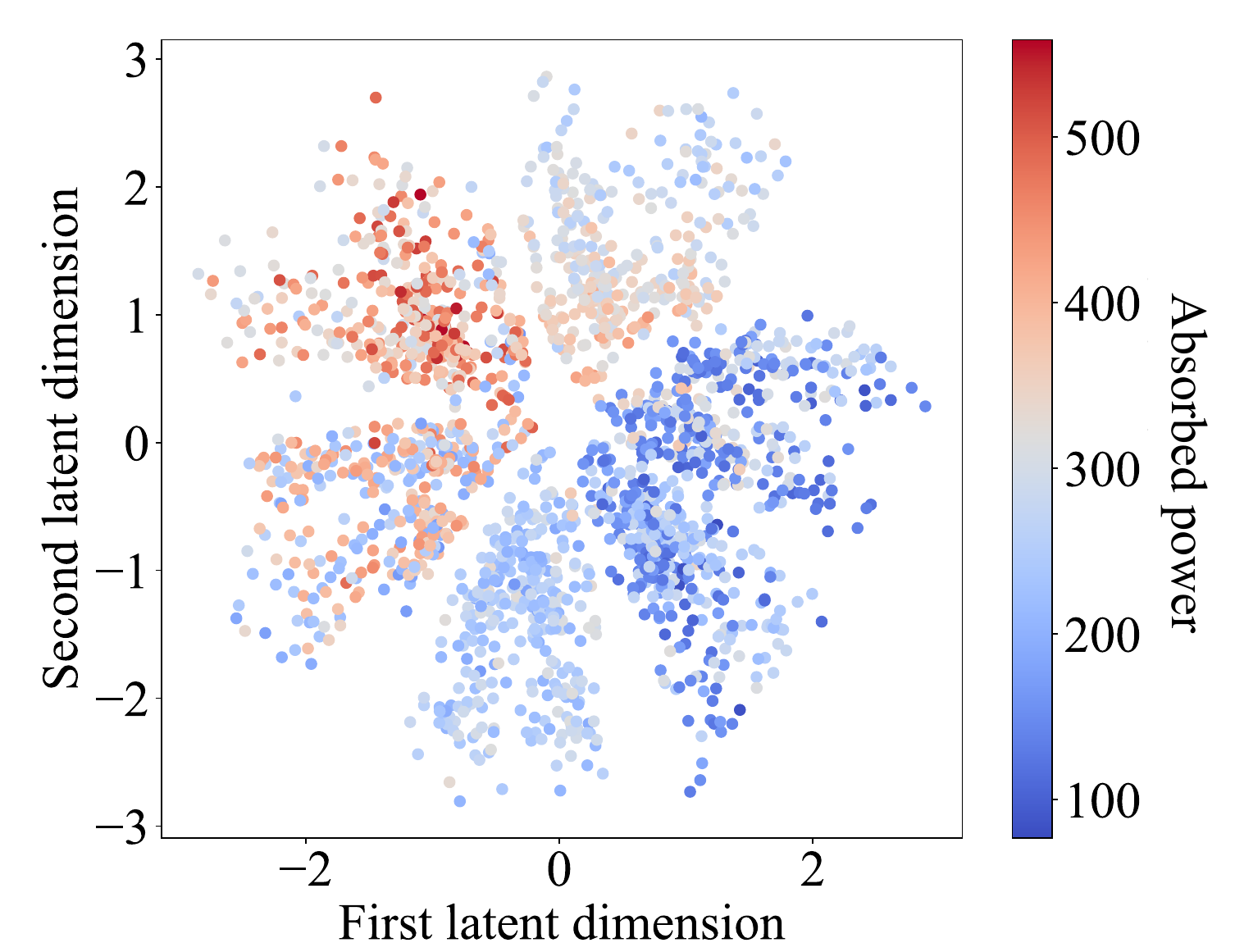}
\caption{\caphead{Correspondence of absorbed power to
the bottom-right-to-upper-left diagonal}}
\label{fig_Latent_Space_Dissipation}
\end{subfigure}
\begin{subfigure}{0.4\textwidth}
\centering
\includegraphics[width=1\textwidth]{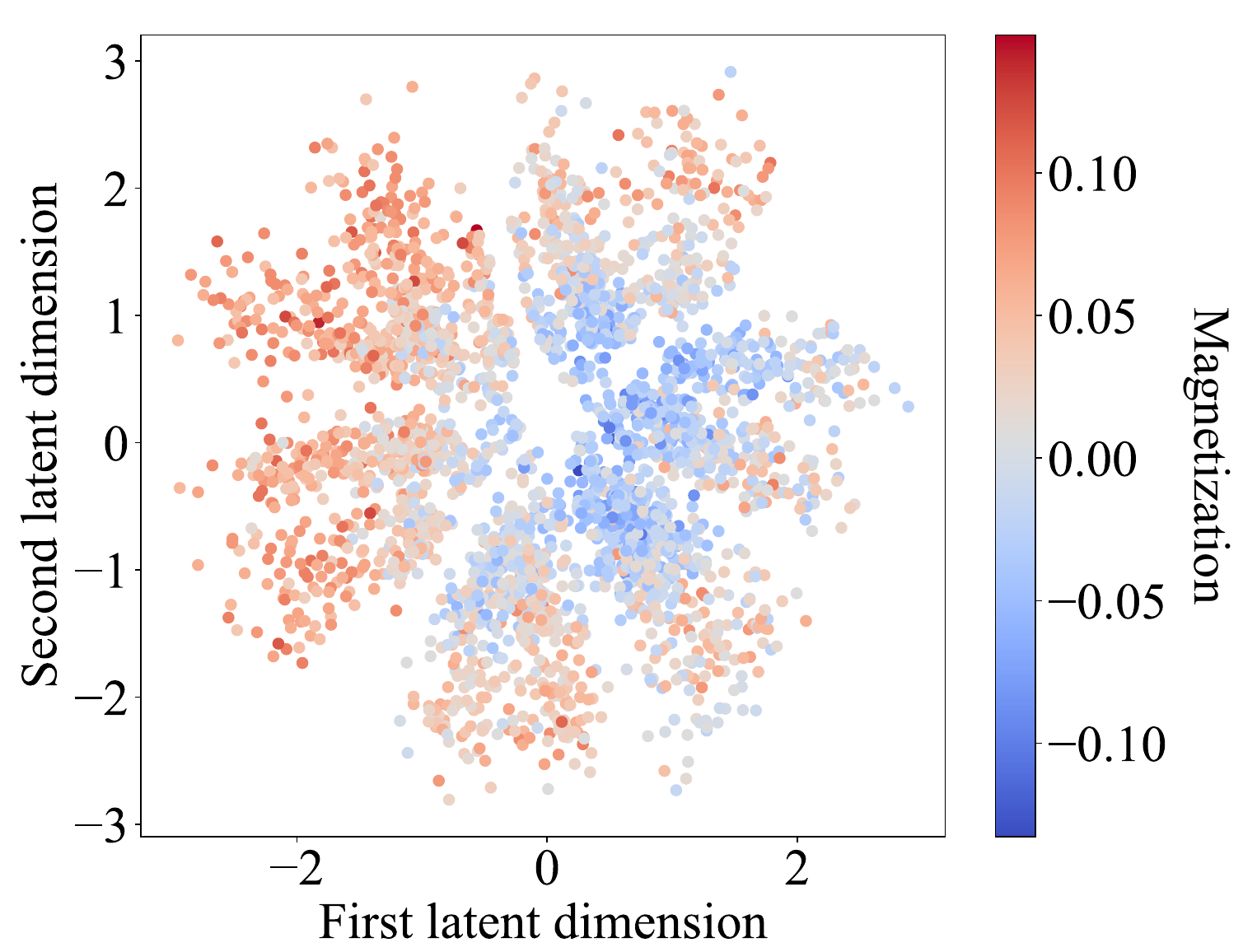}
\caption{\caphead{Correspondence of magnetization to the radial direction.}}
\label{fig_Latent_Space_Magnetization}
\end{subfigure}
\caption{\caphead{Correspondence of latent-space directions
to thermodynamic quantities:}
Each plot depicts the latent space constructed by
a variational autoencoder (VAE).
The VAE trained on the configurations assumed by a spin glass 
during its training with fields $A$, $B$, and $C$.
We have color-coded each plot to highlight
how a thermodynamic property changes 
along some direction.
According to Fig.~\ref{fig_Latent_Space_Dissipation}, 
the absorbed power grows
from the bottom righthand corner to the upper lefthand corner.
According to Fig.~\ref{fig_Latent_Space_Magnetization}, 
the magnetization grows along the radial direction.}
\label{fig_Latent_Space_Visualize}
\end{figure}

Figure~\ref{fig_Latent_Space_Visualize} shows 
two visualizations of the latent space.
Each visualization spotlights a correlation between 
a latent-space direction and a thermodynamic variable.
In Fig.~\ref{fig_Latent_Space_Dissipation},
blue dots represent configurations in which
the spin glass absorbs little work.
Red dots highlight high-absorbed-power configurations.
The dots change from blue to red along the diagonal
from the lower right-hand corner to the upper left-hand.
Hence a point's distance along the diagonal
correlates with the absorbed power.

In Fig.~\ref{fig_Latent_Space_Magnetization},
blue represents low magnetizations,
and red represents high.
Blue dots cluster near the latent space's center,
and red dots occupy the outskirts.
Hence magnetization correlates with a dot's radial coordinate.
Magnetization correlates, to some extent, also with 
distance along the bottom-right-to-upper-left diagonal.
After all, magnetization is related to the absorbed power.

In summary, the diagonal corresponds roughly to the absorbed power,
and the radial direction corresponds roughly to magnetization.
The directions are nonorthogonal, suggesting 
a nonlinear relationship between the thermodynamic variables.
We leave the parameterization of the relationship,
and the possible decoding of other latent-space directions into
new thermodynamic variables, for future work.

%
%
%
\subsection{Feasibility}
\label{sec_Feasibility}

Applying our toolkit might appear impractical,
since microstates must be inputted into the NN.
Measuring a many-body system's microstate may daunt experimentalists.
Yet the use of microstates hinders our proposal little, for three reasons.

First, microstates can be calculated in numerical simulations,
which inform experiments.
Second, many key properties of many-body microstates 
have been measured experimentally.
High-speed imaging has been used to monitor soap bubbles' positions~\cite{Mukherji_19_Strength}
and colloidal suspensions~\cite{Cheng_11_Imaging}.
Similarly wielded tools, such as high magnification, have advanced
active-matter~\cite{Sanchez_12_Spontaneous}
and gene-expression~\cite{Lonsdale_13_Genotype} studies.

One might worry that the full microstate
cannot be measured accurately or precisely.
Soap bubbles' positions can be measured with finite precision,
and other microscopic properties might be inaccessible.
But, third, some bottleneck NNs denoise their inputs~\cite{Vincent_08_Extracting,Goodfellow_16_Deep}:
The NNs learn the distribution from which samples are generated ideally,
not systematic errors.
Denoising by VAEs is less established but is progressing~\cite{Im_15_Denoising}.


Furthermore, one might wonder whether 
our study requires deep learning.
Could simpler algorithms detect and measure many-body learning
as sensitively?
Appendix~\ref{sec_Justify_ML} addresses this question.
We compare the VAE with simpler competitors
that perform unsupervised learning:
a single-layer linear NN,
related to principal-component analysis (PCA)~\cite{Bourland_88_Auto},
and a clustering algorithm.
The VAE outperforms both competitors.

\subsection{Opportunities}
\label{sec_Opportunities}

Several opportunities emerge from 
this combination of statistical mechanical learning
and bottleneck NNs.
First, our toolkit may resolve open problems in the field
of statistical mechanical learning.
One example concerns the soap-bubble raft in~\cite{Mukherji_19_Strength}.
Experimentalists trained a raft of soap bubbles with
an amplitude-$\gamma_{\rm t}$ strain.
The soap bubbles' positions were tracked,
and variances in positions were calculated.
No such measures distinguished trained rafts
from untrained rafts;
only stressing the raft and reading out the strain could~\cite{Mukherji_19_Strength,Miller_19_Raft}.
Bottleneck NNs may reveal what microscopic properties distinguish
trained from untrained rafts.

Similarly, representation learning might facilitate 
the detection of active matter.
Self-organization is detected now through 
simple, large-scale, easily visible signals~\cite{Heylighen_02_Science}.
Bottleneck NNs could identify patterns
invisible in thermodynamic measures.

Second, our framework calls for extensions to quantum systems.
Far-from-equilibrium many-body systems have been realized with 
many quantum platforms, including ultracold atoms~\cite{Langen_15_Ultracold},
trapped ions~\cite{Friis_18_Observation,Smith_16_Many}, 
and nitrogen vacancy centers~\cite{Kucsko_18_Critical}.
Applications to memories have been proposed~\cite{Abanin_19_Colloquium,Turner_18_Weak}.
Yet quantum memories that remember 
\emph{particular coherent states} have been focused on.
The learning \emph{of strong drives} by quantum many-body systems
calls for exploration,
as the learning of strong drives by polymers, soap bubbles, etc.
has proved so productive in classical statistical mechanics.
Our framework can guide this exploration.

Third, we identified a parallel between representation learning and
statistical mechanics. 
The parallel enabled us to use representation learning
to gain insight into statistical mechanics.
Recent developments in information-theoretic far-from-equilibrium statistical mechanics 
(e.g.,~\cite{Still_12_Thermodynamics,Parrondo_15_Thermodynamics,Crutchfield_17_Origins,Kolchinsky_17_Dependence})
might, in turn, shed new light on representation learning.

Fourth, the mutual information between configuration and drive
can be calculated as a function of time.
Let $p(x, y)$ denote the joint probability that
the configuration $X = x$ and the drive $Y = y$.
Let $p(x)  :=  \sum_y  p(x, y)$ and 
$p(y)  :=  \sum_x  p(x, y)$ denote the marginal distributions.
The mutual information quantifies the information 
about the drive in the configuration and vice versa:
$I(X; Y)  
=  \sum_{x, y}  p(x, y)
\log  \left(  \frac{ p(x, y) }{ p(x)  p(y) }  \right)$.
The mutual information is expected to grow
as the many-body system learns.
Estimating $I(X; Y)$ proved difficult due to undersampling;
hence our use of the MAP-estimate score (Sec.~\ref{sec_Classify}),
a cousin of the mutual information (App.~\ref{app_MAP}).
This work motivates the development of 
techniques for estimating $I(X; Y)$ from little data.

Such techniques could be complemented by
a sampling strategy based on our VAE, fifth.
The VAE populates latent space, analogous to the space of macrostates, 
as in Fig.~\ref{fig_Latent_Space}.
Consider choosing an unpopulated point,
analogous to an unfamiliar macrostate,
and having the VAE decompress the point.
The VAE will construct a configuration.
Such configurations could improve $p(x, y)$ estimates
and so $I(X; Y)$ estimates.
Rough initial studies suggest that the constructed configurations
resemble the true samples that they should mimic.

Sixth, given $I(X; Y)$, one can benchmark the many-body system
against the \emph{information curve}~\cite{Tishby_00_Information}.
The information curve quantifies the tradeoff in representation learning:
Recall the general bottleneck NN described in the introduction.
The NN compresses $X$ into $Z$,
then decompresses $Z$ into $Y$ [Fig.~\ref{fig_VAE_SM_Parallel}(a)].
The more the NN compresses $X$,
the less space $Z$ requires.
Hence shrinking $I(X; Y)$ is desirable.
Yet $Z$ must carry enough information about $X$
to generate an accurate $Y$ prediction $\hat{Y}$.
Hence $I(Z; \hat{Y})$ should be large.
One can tune the mutual informations' relative importance,
using a parameter $\beta$.
One chooses a $\beta \in [0, 1]$, then maximizes the objective function
$I(Z; \hat{Y} )  -  \beta I (Z; X)$.
This strategy is called the \emph{information bottleneck}~\cite{Tishby_99_Information}.
Consider varying $\beta$.
At each $\beta$ value, the optimal $I(X; Y)$ can be 
plotted against the optimal $I(X; Z)$.
The resulting \emph{information curve} represents an ideal:
Physical systems can reach the points inside the curve,
not points outside.
Consider plotting a many-body system's 
$\LParen  I(X; Z),  I(X; Y)  \RParen$ as a point.
The point's distance from the information curve
will quantify how close the many-body system approaches to the ideal.

Seventh, we partially decoded the VAE's latent space 
in terms of thermodynamic variables (Sec.~\ref{sec_Decode_Latent}).
Further analysis merits exploration.
Convention biases thermodynamicists toward measuring
volume, magnetization, heat, work, etc.
The VAE might identify new macroscopic variables
better-suited to far-from-equilibrium statistical mechanics,
or hidden nonlinear relationships amongst thermodynamic variables.
A bottleneck NN could uncover new theoretical physics,
as discussed in, e.g.,~\cite{Carleo_19_Machine,Wu_19_Toward,Iten_20_Discovering}.

%
%
\begin{acknowledgments}
The authors thank Alexander Alemi, Isaac Chuang, Emine Kucukbenli, Nick Litombe, Seth Lloyd, Julia Steinberg, Tailin Wu, and Susanne Yelin for useful discussions.
WZ is supported by ARO Grant W911NF-18-1-0101; 
the Gordon and Betty Moore Foundation Grant, under No. GBMF4343;
and the Henry W. Kendall (1955) Fellowship Fund.
JMG is funded by the AFOSR, under Grant FA9950-17-1-0136.
SM was supported partially by the Moore Foundation, 
via the Physics of Living Systems Fellowship.
This material is based upon work supported by, or in part by, the Air Force
Office of Scientific Research, under award number FA9550-19-1-0411.
JLE has been funded by the Air Force Office of Scientific Research grant FA9550-17-1-0136 and by the James S. McDonnell Foundation Scholar Grant 220020476.
NYH is grateful for an NSF grant for the Institute for Theoretical Atomic, Molecular, and Optical Physics at Harvard University and the Smithsonian Astrophysical Observatory.
NYH also thanks CQIQC at the University of Toronto, the Fields Institute, and Caltech's Institute for Quantum Information and Matter (NSF Grant PHY-1733907) for their hospitality during the development of this paper.
\end{acknowledgments}

\begin{appendices}

\onecolumngrid

\renewcommand{\thesection}{\Alph{section}}
\renewcommand{\thesubsection}{\Alph{section} \arabic{subsection}}
\renewcommand{\thesubsubsection}{\Alph{section} \arabic{subsection} \roman{subsubsection}}

\makeatletter\@addtoreset{equation}{section}
\def\theequation{\thesection\arabic{equation}}

\section{Details about the variational autoencoder}
\label{sec_NN_Details}


We briefly motivate and review VAEs,
then describe the VAE applied in the main text.
Further background about VAEs can be found in~\cite{Kingma_13_Auto,JR_14_Stochastic,Doersch_16_Tutorial}.
We denote vectors with boldface in this appendix.

Denote by $\mathbf{X}$ data that has a probability 
$p_{ \bm{\theta} }(\mathbf{x})$
of assuming the value $\mathbf{x}$.
$\bm{\theta}$ denotes a parameter,
and $p_{ \bm{\theta} } (\mathbf{x})$ is called the \emph{evidence}.
We do not know the form of $p_{ \bm{\theta} }(\mathbf{x})$,
when using representation learning.
We model $p_{ \bm{\theta} }(\mathbf{x})$ by identifying 
latent variables $\mathbf{Z}$ 
that assume the possible values $\mathbf{z}$.
Let $p_{ \bm{\theta} }(\mathbf{x} | \mathbf{z})$ 
denote the conditional probability that
$\mathbf{X} = \mathbf{x}$, given that $\mathbf{Z} = \mathbf{z}$.
We model the evidence, using the latent variables, with
\begin{align}
 p_{ \bm{\theta} }(\mathbf{x}) 
 = \int d\mathbf{z} \; 
 p_{ \bm{\theta} }( \mathbf{x} | \mathbf{z} ) 
 p( \mathbf{z} ) .
\end{align}

$p_{ \bm{\theta} }(\mathbf{x} | \mathbf{z} )$ 
can be related to the posterior distribution 
$p_{ \bm{\theta} } (\mathbf{z} | \mathbf{x})$.
The posterior is the probability that, if 
$\mathbf{X} = \mathbf{x}$, then $\mathbf{Z} = \mathbf{z}$. By Bayes' rule,
$p_{ \bm{\theta} }( \mathbf{z} | \mathbf{x} ) 
= p_{ \bm{\theta} }( \mathbf{x} | \mathbf{z} )
p( \mathbf{z} )  /  p_{ \bm{\theta} }( \mathbf{x} )$. 
Calculating the posterior is usually impractical,
as $p_{ \bm{\theta} }(\mathbf{x})$ 
is typically intractable (cannot be calculated analytically).
Hence we approximate the posterior with a variational model 
$q_{ \bm{\phi} }( \mathbf{z} | \mathbf{x} )$.
The optimization parameter $\bm{ \phi }$ denotes 
the NN's weights and biases.

The approximation introduces an inference error,
quantified with the Kullback-Leibler (KL) divergence.
Let $P(\mathbf{u})$ and $Q(\mathbf{u})$ denote distributions over
the possible values $\mathbf{u}$ of a variable.
The KL divergence quantifies the distance between the distributions:
\begin{align}
   D_\KL \LParen P(\mathbf{u}) || Q(\mathbf{u}) \RParen
   & :=  \mathbb{E}_{ P(\mathbf{u}) }
        \left[  \ln P( \mathbf{u} ) \right]
        -  \mathbb{E}_{ P(\mathbf{u}) }
        \left[ \ln  Q(\mathbf{u})  \right] \\
   \label{eq_D_KL_Nonneg}
   & \geq  0 .
\end{align}
We denote by $\mathbb{E}_{ P(\mathbf{u}) } [ f( \mathbf{u} ) ]$
the expectation value of a function $f(\mathbf{u})$.
Operationally, the KL divergence equals the maximal efficiency 
with which the distributions can be distinguished, 
on average, in a binary hypothesis test.
We quantify our inference error with the KL divergence between
the variational model and the posterior,
$D_\KL \LParen  
q_{ \bm{\phi} }( \mathbf{z} | \mathbf{x} ) || 
                         p_{ \bm{\theta} }( \mathbf{z} | \mathbf{x} )  \RParen$. 

Recall that we wish to estimate $p_{\bm \theta} ( \mathbf{x} )$:
An accurate estimate lets us predict $\mathbf{x}$ accurately.
We wish also to estimate the latent posterior distribution, 
$q_{\bm \phi} (\mathbf{z} | \mathbf{x} )$.
We therefore write out the KL divergence's form,
apply Bayes' rule to rewrite the $p_{ \bm{\theta} }( \mathbf{z} | \mathbf{x} )$,
rearrange terms,
and repackage terms into a new KL divergence:
\begin{align}
   \label{eq_Log_Like_1}
   \ln p_{ \bm{\theta} } ( \mathbf{x} )
   =  D_\KL \LParen  q_{\bm \phi} ( \mathbf{z} | \mathbf{x} )
                                                     || p_{\bm \theta} ( \mathbf{z} | \mathbf{x} )
                   \RParen
   +  \mathbb{E}_{ q_{\bm \phi} ( \mathbf{z} | \mathbf{x} )}
   \left[ \ln p_{\bm \theta} ( \mathbf{x} | \mathbf{z} )  \right]
   -  D_\KL  \LParen  q_{\bm \phi} ( \mathbf{z} | \mathbf{x} )
                                                       || p ( \mathbf{z} )  \RParen .
\end{align}
The penultimate term encodes our first goal;
and the final term, our second goal.

Recall that the KL divergence is nonnegative.
The sum of the final two terms therefore lower-bounds the log-likelihood,
$\ln p_{\bm \theta}( \mathbf{x} )$.
$\mathbf{x}$ denotes the event observed,
$\bm{\theta}$ denotes a possible cause,
and $p_{\bm \theta}$ denotes the likelihood that 
$\bm{\theta}$ caused $\mathbf{x}$.
Maximizing each side of Eq.~\eqref{eq_Log_Like_1},
and invoking Ineq.~\eqref{eq_D_KL_Nonneg}, yields
\begin{align}
   \label{eq_Log_Like_2}
   \max_\theta  \left\{   \ln p_{ \bm{\theta} } ( \mathbf{x} )   \right\}
   \geq \max_\theta  \left\{
   \mathbb{E}_{ q_{\bm \phi} ( \mathbf{z} | \mathbf{x} )}
   \left[ \ln p_{\bm \theta} ( \mathbf{x} | \mathbf{z} )  \right]
   -  D_\KL  \LParen  q_{\bm \phi} ( \mathbf{z} | \mathbf{x} )
                                                       || p ( \mathbf{z} )  \RParen 
   \right\} .
\end{align}
The RHS is called the \textit{evidence lower bound} (ELBO). 

A VAE is a neural network that implements the ELBO.
$q_{\bm \phi} ( \mathbf{z} | \mathbf{x} )$ encodes the input $\mathbf{X}$, 
and $p_{\bm \theta} ( \mathbf{x} | \mathbf{z} )$ decodes.
The VAE has the cost function
\begin{equation}
   \label{eq_VAE_Cost}
   \mathcal{L}_{\text{VAE}} 
   := \mathbb{E}_{  p_{\rm emp}( \mathbf{x} )  }
   \left[ \mathbb{E}_{  q_{\bm \phi} ( \mathbf{z} | \mathbf{x} )  }
   \left[  \ln p_{\bm \theta} ( \mathbf{x} | \mathbf{z} )  \right] 
   -  D_\KL  \LParen  q_{\bm \phi} ( \mathbf{z} | \mathbf{x} ) 
                                                        \| p( \mathbf{z} )  \RParen \right].
\end{equation}
$p_{\rm emp}( \mathbf{x} )$ denotes the distribution inferred from the empirical dataset.
Given input values $\mathbf{x}$, the VAE generates a latent distribution 
$q_{\bm \phi} ( \mathbf{z} | \mathbf{x} ) 
= \mathcal{N}( \bm{\mu}_{ \mathbf{z} | \mathbf{x} }, 
                                           \bm{\Sigma}_{ \mathbf{z} | \mathbf{x} } )$.
We denote by $\mathcal{N} ( \bm{\mu}, \bm{\Sigma} )$
the standard multivariate normal distribution
whose vector of means is $\bm{\mu}$ 
and whose covariance matrix is $\bm{\Sigma}$.
Neural-network layers parameterize the VAE's
$\bm{\mu}_{ \mathbf{z} | \mathbf{x} }$ and 
$\bm{\Sigma}_{ \mathbf{z} | \mathbf{x} }$.
Latent vectors are sampled according to 
$q_{\bm \phi}( \mathbf{z} | \mathbf{x} )$,
then decoded into outputs distributed according to
$p_{\bm \theta} ( \mathbf{x} | \mathbf{z} ) 
= \mathcal{N}( \bm{\mu}_{ \mathbf{x} | \mathbf{z} }, 
                        \sigma^2_{ \mathbf{x} | \mathbf{z} }  \id  )$.
Neural-network layers parameterize the mean vector 
$\bm{\mu}_{ \mathbf{x} | \mathbf{z} }$.
The variance $\sigma^2_{ \mathbf{x} | \mathbf{z} }$ is a hyperparameter.

A VAE with the following architecture produced the results in the main text.
The style was borrowed from~\cite{Hafner_18_Building}.
Two fully connected $200$-neuron hidden layers
process the input data.
One fully connected two-neuron hidden layer parameterizes each of
$\bm{\mu}_{ \mathbf{z} | \mathbf{x} }$ and 
$\bm{\Sigma}_{ \mathbf{z} | \mathbf{x} }$.
Two fully connected $200$-neuron hidden layers process the latent variables.
An output layer reads off the outputs. 
We choose $\sigma^2_{ \mathbf{x} | \mathbf{z} }=1$ 
and use Rectified Linear Unit (ReLU) activations
for all hidden layers.

\section{Distinction between robust learning \\
and two superficially similar behaviors}
\label{sec_Not_Enslaved_Or_Frozen}


Learning contrasts with two other behaviors
that the spin glass could exhibit,
entraining to the field and near-freezing.

\subsection{Entraining to the field}
\label{sec_Enslave}

Imagine that most spins align with any field $A$. 
The configuration reflects the field
as silly putty reflects the print of a thumb
pressing on the silly putty.
Smoothing the silly putty's surface wipes the thumbprint off.
Similarly, applying a field $B \neq A$ to the spin glass
wipes the signature of $A$ from the configuration.
From the perspective of the end of the application of $B$,
the spin glass has not learned $A$.
The spin glass lacks a long-term memory of the field;
the field is encoded in no robust, deep properties of the configuration.

We can distinguish learning from entraining
by calculating the percentage of the spins 
that align with the field at the end of training.
If the spins obeyed the field, 100\% would align.
If the spins ignored the field, $50\%$ would align, on average.
Hence the spin glass's entraining is quantified with
\begin{align}
   2 ( \text{Percentage of spins aligned with the field} )  - 100 .
\end{align}
(This measure does not apply to alignment percentages $< 50$,
which are unlikely to be realized.)

\begin{figure}[hbt]
\centering
\includegraphics[width=.45\textwidth, clip=true]{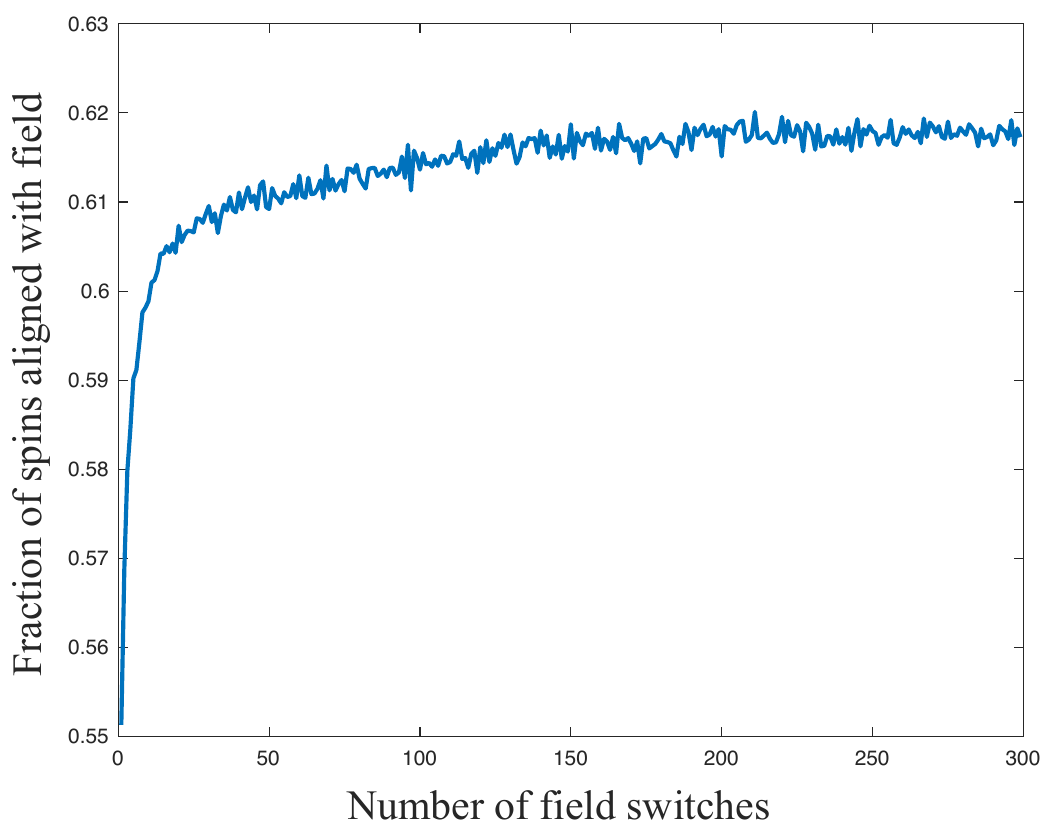}
\caption{\caphead{Fraction of the spins aligned with the field,
as a function of time:}
If a fraction $\approx 1$ of the spins align, the spin glass resembles silly putty,
which shallowly reflects the print of a thumb that presses on it.
Robust learning stores information deep in a system's structure.}
\label{fig_Frac_Aligned}
\end{figure}

Figure~\ref{fig_Frac_Aligned} shows data collected 
about the spin glass in the good-learning regime (Sec.~\ref{sec_Spin_Glass}).
The number of aligned spins is plotted against
the amount $t$ of time for which the spin glass has trained.
After the application of one field, 55\% of the spins align with the field.
At the end of training, 62\% align. 
Hence the spins' entraining grows from 10\% to 24\%.
Growth is expected, as the spin glass learns the training drive.
But $24\%$ is an order of magnitude less than $100\%$, 
so the spin glass is not entrained to the field.

\subsection{Near-freezing}
\label{sec_Near_Freeze}

Suppose that the spin glass is nearly frozen.
Most spins cannot flip, but a few jiggle under most fields.
The spin glass does not learn any field effectively,
being mostly immobile.
But the few flippable spins reflect the field.
A bottleneck NN could guess the field from those few spins.
The NN's low loss function would induce a false positive, 
leading us to believe that the spin glass had learned.


We can avoid false positives by measuring two properties.
First, we measure the percentage of the spins
that antialign with the field.
If the percentage consistently $\gg 0$,
many of the spins are not frozen.
Figure~\ref{fig_Frac_Aligned} confirms that many are not.

\begin{figure}[hbt]
\centering
\includegraphics[width=.5\textwidth, clip=true]{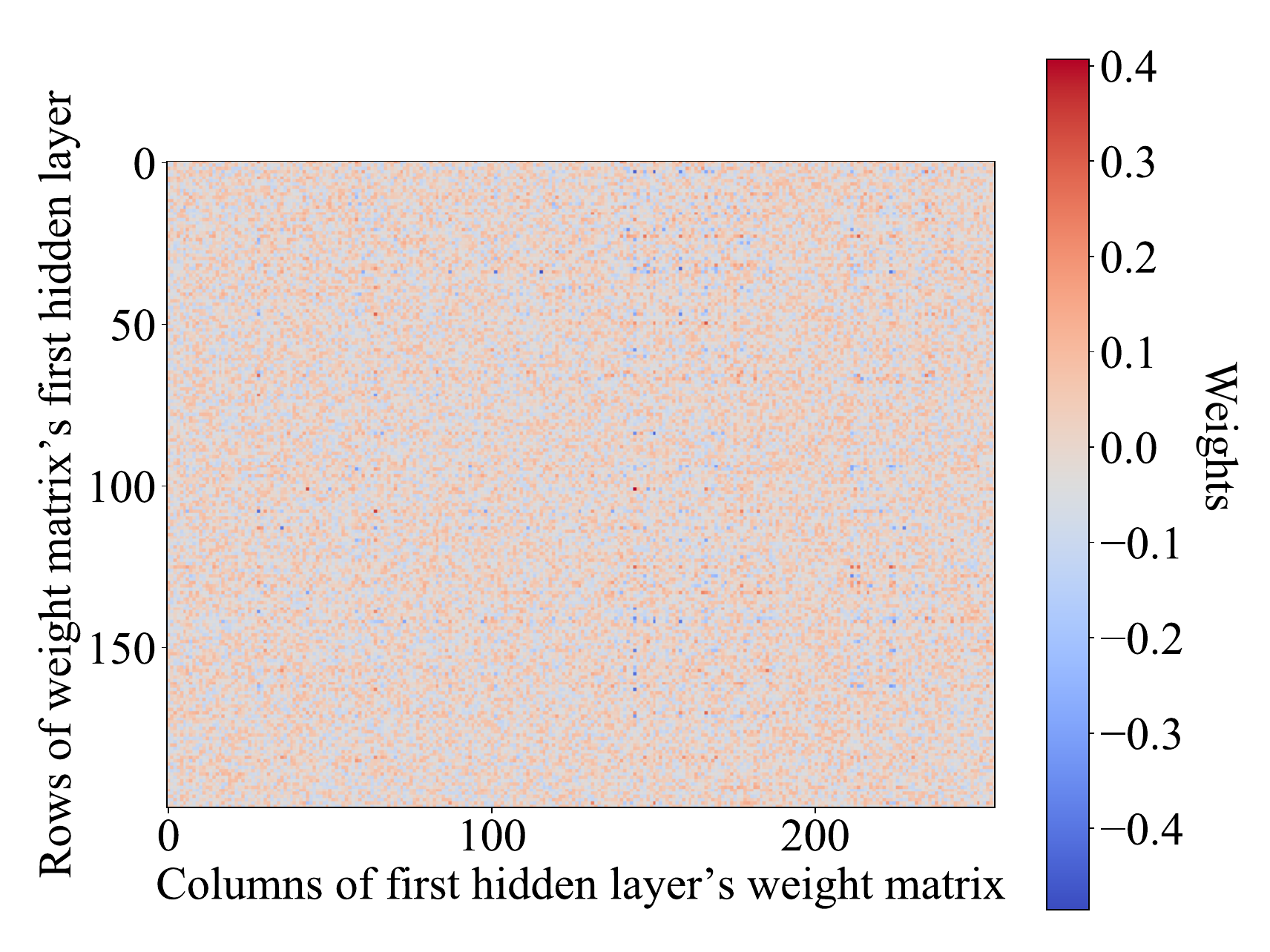}
\caption{\caphead{How much information about each spin 
the variational autoencoder compresses:}
This figure represents the first hidden layer's weight matrix.
The weight matrix transforms the input layer, which consists of 256 neurons,
into the first hidden layer, which consists of 200 neurons.
The matrix's elements are replaced with colors.
Each vertical line corresponds to one spin.
The farther leftward a stripe, the lesser the spin's field energy 
[Eq.~\eqref{eq_Hamiltonian_j}].
} 
\label{fig_Which_Spins_Used_By_AE_b}
\end{figure}

Second, we check that the VAE compresses information about
spins that have many different field energies $A_j(t) s_j$
[Eq.~\eqref{eq_Hamiltonian_j}].
We illustrate with the protocol used to generate Fig.~\ref{fig_Latent_Space_Visualize}:
We trained the spin glass on a drive $\{A, B, C\}$
in each of many trials.
On the end-of-trial configurations, the VAE was trained.

A configuration is represented in the VAE's input layer, a column vector.
A weight matrix transforms the input layer
into the first hidden layer, another column vector.
The weight matrix is depicted in Fig.~\ref{fig_Which_Spins_Used_By_AE_b}.
The matrix's numerical entries have been replaced with colors.
Each vertical stripe corresponds to one spin.
The farther leftward a stripe, the lesser the spin's field energy.
The darker a stripe, the more information about the spin 
the VAE uses when forming $Z$.
The plot is approximately invariant, at a coarse-grained level,
under translations along the horizontal.
(On the order of ten exceptions exist.
These vertical stripes contain several dark dots.
An example appears at $x \approx 150$.
But the number of exceptions is much less than the number of spins:
$\approx 10 \ll 256$.)
Hence the NN uses information about spins of many field energies.
The spins do not separate into
low-field-energy flippable spins
and high-field-energy frozen spins.

\section{Maximum \emph{a posteriori} estimation (MAP estimation)}
\label{app_MAP}



This appendix details the MAP estimation
applied in Sections~\ref{sec_Classify}-\ref{sec_Discriminate}.
MAP estimates help answer the question
``How accurately can the drive be identified from the spin configuration?''
We return to the notation used in the introduction,
denoting the drive by $Y$ and the configuration by $X$.

In information theory, we answer this question using the conditional entropy,
\begin{align}
   \label{eq_Cond_Ent}
   H(Y | X) := - \sum_{x, y} p(x, y)  \log \frac{ p(x, y) }{ p(x) } .
\end{align}
$p(x, y)$ denotes a joint distribution; and $p(x)$, a marginal.
The conditional entropy quantifies
the uncertainty about the drive, given the configuration. 
Equation~\eqref{eq_Cond_Ent} does not refer to
any estimator of $Y$.
Rather, $H(Y | X)$ underlies a bound on the accuracy with which
any estimator can reconstruct the drive from the configuration,
by Fano's inequality.
Estimating $H(Y | X)$ proves difficult, due to undersampling:
An enormous amount of data is needed to estimate the distribution
$p(y | x)$ accurately enough to estimate $H(Y | X)$
(Sec.~\ref{sec_Opportunities}).

Undersampling plagues also the mutual information,
a sister of the conditional entropy:
$I(X; Y)  :=  H(Y) - H(Y | X)$.
The Shannon entropy, $H(Y) := - \sum_y p(y) \log p(y)$,
quantifies the randomness in the drive variable.
The mutual information quantifies 
the information about the drive in the configuration and vice versa.

$H(Y | X)$ and $I(X; Y)$ offer one answer to our question.
Another comes from using MAP estimation 
to predict drives from configurations,
then scoring the predictions.
MAP estimation proceeds as follows.
One approximates the conditional probability distribution $p(y|x)$ 
from the data via any possible strategy.
(We detail one strategy below.)
Let $\tilde{p} (y | x)$ denote the approximation.
Given a configuration $x$, one predicts that it resulted from the drive
\begin{equation}
   \label{eq_How_To_MAP}
    \hat{y} = \arg\max_{y}  \Set{  \tilde{p} ( y | x )  }
\end{equation}
that has the greatest conditional probability.
Equation~\eqref{eq_How_To_MAP} is the MAP estimator.
We use it to map all the configurations $x$
to drive predictions $\hat{y}$.
The frequency with which $\hat{y} = y$ is the estimator's score.

To use the MAP estimator~\eqref{eq_How_To_MAP},
we must approximate the conditional probability distribution $p(y | x)$.
Our approximation suffers from undersampling.
Hence we invoke the map $f(x) = z$ from configurations $x$
to the low-dimensional latent-space variable $z$.
Approximating $p \LParen y | f(x) \RParen$ proves easier than
approximating $p(y | x)$. By Bayes' rule, 
$p \LParen y | f(x) \RParen
=  \frac{  p \LParen f(x) | y \RParen  p(y)  }{  p \LParen f(x) \RParen  }$.
The approximation $\tilde{p} \LParen y | f(x) \RParen$ factors analogously.
We redefine our estimator as
\begin{align}
   \hat{y}
   & =  \arg  \max_y  \Set{  \tilde{p} \LParen y | f(x) \RParen  } 
   =  \arg \max_y  \Set{
   \frac{  \tilde{p} \LParen f(x) | y \RParen  \tilde{p}(y)  }{  
             \tilde{p} \LParen f(x) \RParen  }  }
   =  \arg \max_y  \Set{   \tilde{p} \LParen f(x) | y \RParen  \tilde{p}(y)  } .
\end{align}
The final equality holds because the arg-max over $y$ cannot depend on 
the $y$-independent $\tilde{p} \LParen f(x) \RParen$.
The fields $y$ are chosen uniformly randomly from the drive.
Hence $p(y) \approx \tilde{p}(y)$ is constant, and
\begin{align}
   \hat{y}
   & \approx  \arg  \max_y  \Set{  \tilde{p} \LParen y | f(x) \RParen  } .
\end{align}
This MAP estimate equals the maximum-likelihood estimate.
Generally, MAP estimation with a uniform prior
amounts to maximum-likelihood estimation.
We use only uniform priors.
Other applications of our toolkit, however,
can benefit from alternative priors, if extra information is available.
Hence we present the MAP generalization of maximum-likelihood estimation.
A Gaussian distribution approximates $p \LParen y | f(x) \RParen$ well,
so $\hat{y}$ can be approximated easily.

\section{Memory capacity attributed to the many-body system \\
by the absorbed power}
\label{app_Capacity_Work}


In Sec.~\ref{sec_Capacity}, we compared the memory capacity
registered by the VAE
to the capacity registered by the absorbed power.
The study involved MAP estimation on 
drives of 40 fields selected from 50 fields.
The choice of 50 is explained here:
Fifty fields exceed the spin-glass capacity
registered by the absorbed power.

Recall how memory has been detected thermodynamically~\cite{Gold_19_Self}:
Let a many-body system be trained with
a drive that includes a field $A$.
Consider testing the system, afterward, 
with an unfamiliar field $B$, and then with $A$.
Suppose that the absorbed power jumps substantially when $B$ is applied
and less when $A$ is reapplied.
The many-body system identifies $B$ as novel
and remembers $A$,
according to the absorbed power.

We sharpen this analysis.
First, we divide the trial into time windows.
During each time window, the field switches 10 times.
(The 10 eliminates artificial noise and is not critical.
Our qualitative results are robust with respect to changes in such details.)
We measure the absorbed power at the end of each time window
and at the start of the subsequent window.
We define ``the absorbed power jumps substantially'' as
``the absorbed power jumps, on average over trials, 
by much more than the noise
(by much more than the absorbed power fluctuates across a trial)'':
\begin{align}
   \label{eq_Capacity_Condn_Work}
   & \langle ( \text{Power absorbed at start of later time window} ) 
   - ( \text{Power absorbed at end of preceding time window} )
   \rangle_{ \text{trials} } 
   \\ \nonumber & 
   \gg  \text{Standard deviation in }
   [ ( \text{Power absorbed at start of later window} )
   -  ( \text{Power absorbed at end of preceding window} ) ] .
\end{align}
Consider including only a few fields in the training drive, 
then growing the drive in later trials.
The drive will tax the spin glass's memory until exceeding the capacity.
The LHS of~\eqref{eq_Capacity_Condn_Work}
will come to about equal the RHS.

\begin{figure}[hbt]
\centering
\includegraphics[width=.45\textwidth, clip=true]{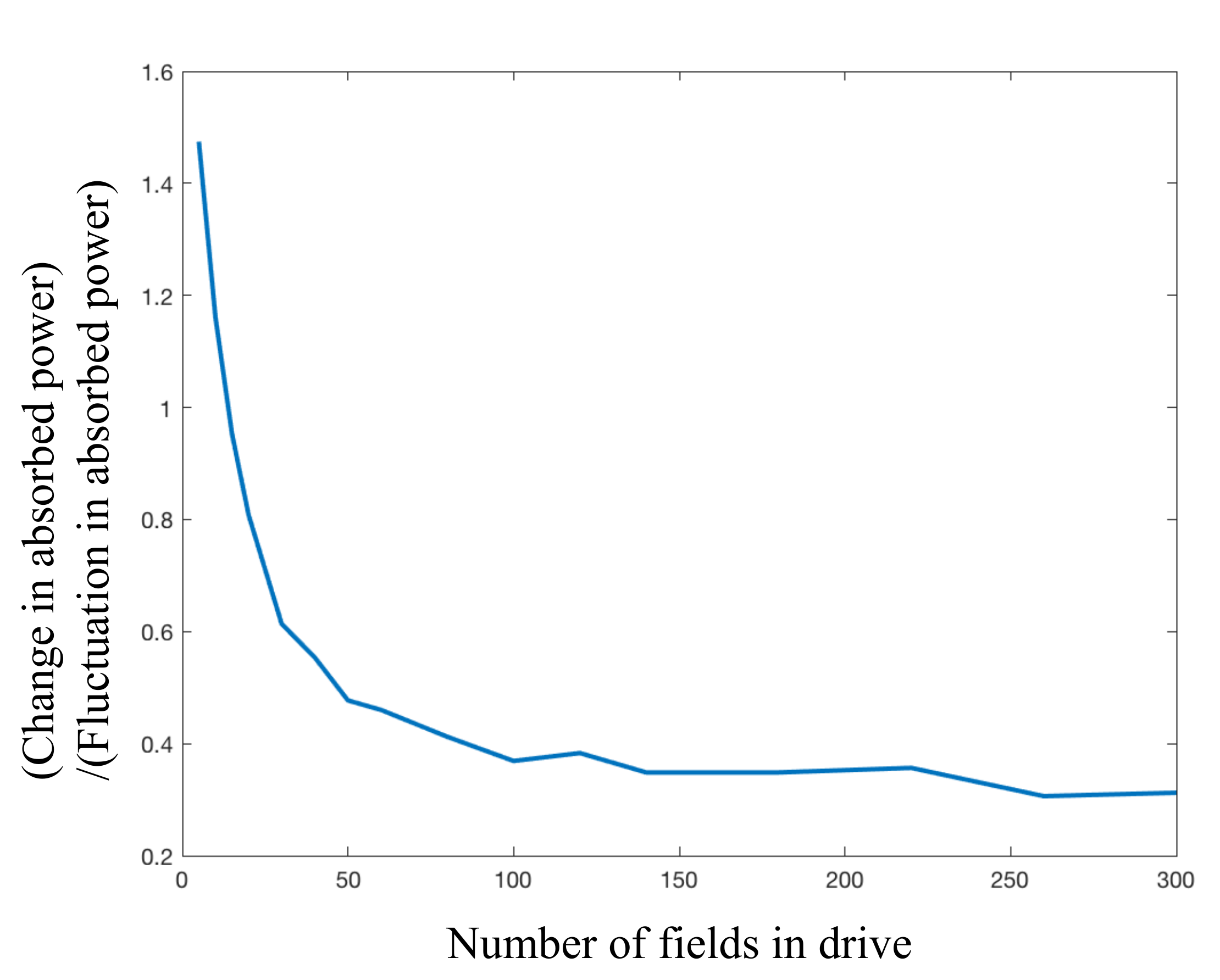}
\caption{\caphead{Estimate of memory capacity by absorbed power:}
A many-body system reaches its capacity, 
according to the absorbed power, when
[left-hand side of Ineq.~\eqref{eq_Capacity_Condn_Work}]
/ (right-hand side) $\approx 1$.
The curve $\approx 1$, and a 256-spin glass reaches its capacity,
when the training drive contains $\approx 15$ fields.}
\label{fig_Capacity_Work}
\end{figure}

Figure~\ref{fig_Capacity_Work} illustrates with the spin glass.
On the $x$-axis is the number of fields in the training drive.
On the $y$-axis is the ratio of 
the left-hand side of Ineq.~\eqref{eq_Capacity_Condn_Work}
to the right-hand side (LHS/RHS).
Where LHS/RHS $\approx 1$, the spin glass reaches its capacity.
This spin glass can remember $\approx 15$ fields,
according to the absorbed power.

\section{Justification of use of machine learning}
\label{sec_Justify_ML}


Deep learning is a powerful tool.
Is it necessary for recovering our results?
Could simpler algorithms detect and quantify many-body learning as sensitively?
Comparable simpler algorithms tend not to, we find.
Two competitors suggest themselves:
single-layer linear autoencoders,
related to PCA~\cite{Bourland_88_Auto}, 
and clustering algorithms.
Alternatives include generalized linear models~\cite{Bishop_06_Pattern}
and supervised linear autoencoders.
These models, however, perform supervised learning.
They receive more information than the VAE
and so enjoy an unfair advantage.
We analyze the two comparable competitors sequentially.

\subsection{Comparison with single-layer linear autoencoder}

The linear autoencoder is a single-layer NN.
The input, $X$, undergoes a linear transformation:
$Y = mX + b$.
We compare, as follows, the linear autoencoder's 
detection of field classification 
with the VAE's detection:
We trained the spin glass on a drive in each of 
3,000-5,000 trials.
Ninety percent of the trials were designated as NN-training data;
and 10\%, as NN-testing data.
For each training trial, we identified the spin glass's final configuration.
On these configurations, each NN performed unsupervised learning.
Each NN then received the configuration with which
the spin glass ended a NN-testing trial.
We inferred the field most likely to have produced this configuration,
using MAP estimation.
The fraction of trials in which the NN points to the correct field 
constitutes the NN's score.
On a three-field drive, the linear autoencoder scored 0.771,
while the VAE scored 0.992.
On a five-field drive, the linear autoencoder scored 0.3934,
while the VAE scored 0.829.
Hence the VAE picks up on more of 
the spin glass's ability to classify fields.

\subsection{Comparison with clustering algorithm}

A popular, straightforward-to-apply algorithm is 
\emph{$k$-means clustering}~\cite{Bishop_06_Pattern}.
$k$ refers to a parameter inputted into the algorithm,
the number of clusters expected in the data.
We inputted the number of drives imposed on the spin glass,
in addition to inputting configurations.
The VAE receives just configurations and so less information.
We could level the playing field by automating the choice of $k$,
using the Bayesian information criterion (BIC)~\cite{Bishop_06_Pattern}.
But clustering with the BIC-chosen $k$ 
would perform no better than
clustering performed with the ideal $k$,
and the ideal clustering performs worse than the VAE.

The protocol run on the spin glass is described 
at the beginning of Sec.~\ref{sec_Capacity}.
Five thousand trials were performed.
The configuration occupied by the spin glass 
at the end of each trial was collected.
Splitting the data into testing and training data
did not alter results significantly.
Hence we fed all the configurations,
with the number $k = 5$ of drives,
to the clustering algorithm.
The algorithm partitioned the set of configurations into subsets.
Each subset contained configurations 
likely to have resulted from the same drive.

Clustering algorithms are assessed with the Rand index,
denoted by RI~\cite{Rand_71_Objective}.
The Rand index differs from the MAP-estimation score (Sec.~\ref{sec_Classify}).
How to compare the clustering algorithm
with the VAE, therefore, is ambiguous.
However, the Rand index quantifies 
the percentage of the algorithm's classifications that are correct.
Hence the Rand index and the MAP-estimation score
have similar interpretations, despite their different definitions.

The clustering algorithm's Rand index began at $\text{RI} = 0$, at $t = 0$.
RI rose during the first $\approx 200$ changes of the drive,
then oscillated around 0.125.
Figure~\ref{fig_Capacity} shows the VAE's performance.
The VAE's score rose during the first $\approx 150$ changes of the drive,
then oscillated around $0.450 > 0.125$.
Hence the VAE outperformed the clustering algorithm.

\end{appendices}

%
%
\bibliographystyle{h-physrev}
\bibliography{Noneq_ML_Bib}


\end{document}